\pdfoutput=1
\documentclass[usenatbib, useAMS]{mnras}
\usepackage[]{graphicx}
\usepackage{amssymb}
\usepackage{epstopdf}
\usepackage[T1]{fontenc}
\usepackage{aecompl}

\newcommand{\ramses}{R{\sc amses}~}

\newcommand{\nut}{N{\sc ut}~}
\newcommand{\nutfb}{N{\sc utfb}}
\newcommand{\vlos}{$v_{\rm LOS}$}
\newcommand{\vdisp}{$\sigma_{*}$}
\newcommand{\kms}{km s$^{-1}$}
\newcommand{\degree}{$^{\circ}$}
\newcommand{\msun}{M$_{\sun}$}

\title[Simulated studies of high-z galaxies with HARMONI]{Simulated stellar kinematics studies of high-redshift galaxies with the HARMONI Integral Field Spectrograph}
\author[Kendrew et al.]
{\parbox{\textwidth}{S. Kendrew$^{1}$\thanks{E-mail: sarah.kendrew@physics.ox.ac.uk}, 
 S. Zieleniewski$^{1}$, R.~C.~W. Houghton$^{1}$, N. Thatte$^{1}$, J. Devriendt$^{1}$, M. Tecza$^{1}$,
F.Clarke$^{1}$, K. O'Brien$^{1}$ and B. H\"{a}u{\ss}ler$^{1, 2, 3}$.}\vspace{0.4cm}\\
\parbox{\textwidth}{$^{1}$University of Oxford, Department of Physics, Denys Wilkinson Building, Keble Road, Oxford OX1 3RH, United Kingdom\\
$^{2}$European Southern Observatory, Alonso de Cordova 3107, Vitacura, Santiago, Chile\\
$^{3}$University of Hertfordshire, Hatfield, Herts AL10 9AB, United Kingdom}}

\begin{document}

\date{Submitted xxx 2015}

\pagerange{\pageref{firstpage}--\pageref{lastpage}} \pubyear{2002}

\maketitle

\label{firstpage}

\begin{abstract}
We present a study into the capabilities of integrated and spatially resolved integral field spectroscopy of galaxies at $z=2-4$ with the future HARMONI spectrograph for the European Extremely Large Telescope (E-ELT) using the simulation pipeline, HSIM. We focus particularly on the instrument's capabilities in stellar absorption line integral field spectroscopy, which will allow us to study the stellar kinematics and stellar population characteristics. Such measurements for star-forming and passive galaxies around the peak star formation era will provide a critical insight into the star formation, quenching and mass assembly history of high-z, and thus present-day galaxies. First, we perform a signal-to-noise study for passive galaxies at a range of stellar masses for $z=2-4$, assuming different light profiles; for this population we estimate integrated stellar absorption line spectroscopy with HARMONI will be limited to galaxies with $M_* \gtrsim 10^{10.7}$\msun. Second, we use HSIM to perform a mock observation of a typical star-forming 10$^{10}$~\msun~galaxy at $z=3$ generated from the high-resolution cosmological simulation \nutfb. We demonstrate that the input stellar kinematics of the simulated galaxy can be accurately recovered from the integrated spectrum in a 15-hour observation, using common analysis tools. Whilst spatially resolved spectroscopy is likely to remain out of reach for this particular galaxy, we estimate HARMONI's performance limits in this regime from our findings. This study demonstrates how instrument simulators such as HSIM can be used to quantify instrument performance and study observational biases on kinematics retrieval; and shows the potential of making observational predictions from cosmological simulation output data.
\end{abstract}


\section{Introduction}

How galaxies assemble their mass over cosmic time, through accretion of gas from the intergalactic medium, mergers and star formation, is a major open question in models of galaxy evolution. Numerous mechanisms are known to regulate the growth of galaxies, such as feedback from massive stars, supernovae and active galactic nuclei (AGN), and long gas cooling times. Their relative importance is however poorly understood. Identifying the influence of different physical mechanisms requires the study of galaxies from the local Universe to high redshifts in all their constituents: stars, gas and dark matter. In both observational and theoretical astrophysics, substantial progress has been made in recent years.

The advent of integral field spectroscopy (IFS) has facilitated the study of spatially resolved kinematics of stars and gas in galaxies. Combining the benefits of imaging and spectroscopy via IFS, a number of major surveys of galaxies in the local Universe, such as SAURON~\citep{Bacon2001}, ATLAS$^{3D}$~\citep{Cappellari2011} and CALIFA~\citep{Sanchez2012} have provided a wealth of new information on the internal dynamics galaxies, their star formation histories, gas abundances, stellar population properties, as well as external environmental effects, such as mergers and tidal stripping. At high redshifts, IFS surveys of gas kinematics such as SINS~\citep{ForsterSchreiber2009} have provided evidence for mass assembly through both continuous accretion and mergers. 

Nebular emission lines (e.g. H$\alpha$, H$\beta$, [OIII]) are accessible to present-day instruments such as VLT SINFONI to $z\sim$~2 (e.g. the SINS survey,~\citealt{ForsterSchreiber2009}), and the KMOS-3D survey is expected to make further inroads into the high-z domain~\citep{Wisnioski2014}. Stellar absorption line spectroscopy is however limited by the continuum S/N that can be achieved with modern facilities, to only $z \sim 0.2$~\citep{Deugenio2013, Rodriguez2014}. Studies of stellar populations at higher redshifts rely on integrated (unresolved) properties. Use of CCDs is possible only to $z \lesssim 1.3$ due to the passing of the Balmer break and Ca H+K features beyond 1~\micron; the advent of efficient near-infrared (NIR) detectors with reduced read-out noise are allowing further advances to $z \sim 2$~\citep[e.g.][]{Belli2014a}.

Although the cosmic star formation history is recognised to peak around $1 < z < 2$~\citep{Madau2014}, the most massive galaxies formed the bulk of their stars at earlier times and over short timescales~\citep{Thomas2005}. Contrary to hierarchical expectations, these massive early-type galaxies are already in place with similar masses at $z \lesssim 4$, just 1.6 Gyr after the Big Bang~\citep{Straatman2014, Marsan2015}. To measure the properties of these pristine passive stellar populations via absorption line spectroscopy requires exceptionally deep observations in the range $2< z < 4$; this is where we expect HARMONI on the E-ELT to break substantial new ground. 

In this paper, we examine the capabilities of the HARMONI IFS~\citep{Thatte2014} on the E-ELT for the study of high-z galaxies in the critical period before the peak star formation era, around $2< z < 4$. As a first step, we present S/N calculations for a range of galaxy masses from $2< z < 4$, assuming simple profiles and star formation histories. A more in-depth study is then performed of a $z=3$ galaxy, whose stellar kinematics we attempt to recover with simulated HARMONI observations.  

For the detailed $z=3$ simulations, we use a dedicated instrument simulation pipeline to produce mock observations of a massive star-forming galaxy at z$\sim$3, extracted from cosmological simulations with the adaptive mesh refinement (AMR) hydrodynamical code \ramses\citep{Teyssier2002} terminated at $z=3$. Full details of this simulation are given in Section~\ref{sec:nutfb}. The advantage of this approach is that the input data are the result of early Universe physics modelled forward through cosmic time, rather than a model-dependent extrapolation of low-$z$ data to higher redshifts. As the input physics are accurately known, we can reliably measure how well the galaxy's properties are recovered from our simulated observations. 

In the following sections, we give a brief overview of the HARMONI instrument and the methods developed to predict its performance. In Section~\ref{sec:hsim} we present HSIM, a dedicated modular Python-based pipeline that produces realistic mock integral field spectroscopic observations. Section~\ref{sec:ryan} shows the first component of our simulation results for HARMONI's predicted performance in the study of high-z galaxies, assuming simple stellar population properties and morphological models. We then describe in Section~\ref{sec:siminput} the data and methodology for a proof-of-concept study of simulated HARMONI observations of a z$\sim$3 galaxy extracted from a cosmological simulation. The method by which the simulation data were converted to a spectral cube is shown, and using the galaxy's integrated spectrum we discuss the observational challenges and chosen strategy for the study in Section~\ref{sec:hsims}. The data analysis methods used to study the kinematic properties of a range of input and output data products are described in Section~\ref{sec:data_analysis}, and the results are then presented in Section~\ref{sec:results}. Finally we discuss the implications for HARMONI observations with the E-ELT in Section~\ref{sec:discussion}. Whilst the simulation pipeline was specific to the E-ELT and HARMONI, our conclusions are likely to be applicable to other IFS instruments in the ELT era.

\section{The HARMONI Instrument and Simulator}\label{sec:hsim}

HARMONI~\citep{Thatte2014} is one of two first-light instruments selected for the 39-m E-ELT, due to come online in the mid-2020s. The instrument will provide integral field spectroscopy from visible to NIR wavelengths ($\sim$0.47 - 2.45~\micron) at a range of spatial scales and spectral resolutions. Operating close to the diffraction limit, it will provide unprecedented gains in sensitivity and spatial resolution. Its design can adapt to several adaptive optics (AO) modes offered by the telescope, such as Single Conjugate AO (SCAO), Laser Tomographic AO (LTAO), as well as seeing limited observations. 

As part of the HARMONI design study, an instrument simulator was developed to investigate the scientific performance of the instrument and test design trade-offs~\citep{Zieleniewski2014, Zieleniewski2015}. The simulator performs an ``observation'' in software of an input spectral cube using the following steps:

\begin{enumerate}
    \item the spectral axis is convolved with a Gaussian instrument line profile of the required width;
    \item AO point spread functions (PSFs) are generated for each wavelength channel and convolved with the image;
    \item atmospheric differential refraction is added along one spatial axis;
    \item atmospheric and telescope radiance and transmission are added to the spectrum;
    \item noise is calculated based on best-estimate detector characteristics;
    \item the output as-observed cube (data and variance) is written out.
\end{enumerate}

\begin{figure}
    \includegraphics[width=9cm]{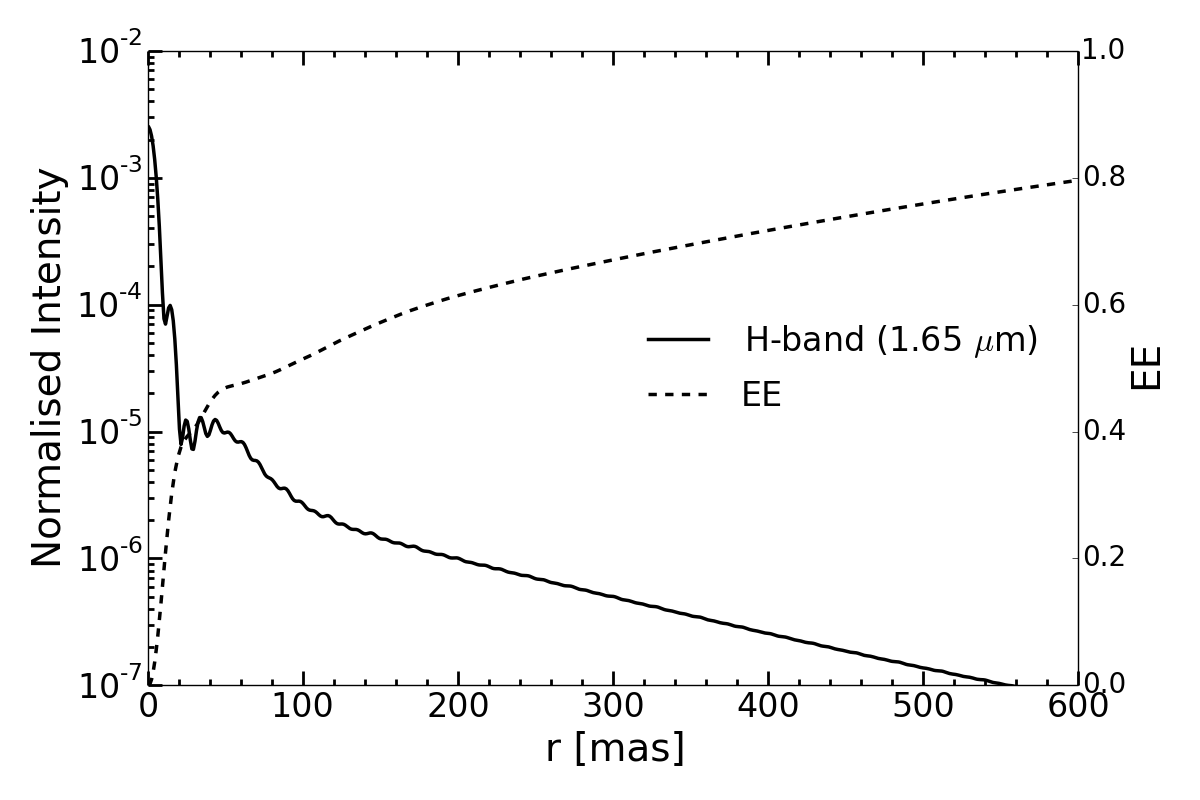}
    \caption{A slice through the simulated LTAO-corrected HARMONI PSF (solid line, left axis) in the NIR H-band (Strehl ratio $\sim$0.35), as used in HSIM, alongside the cumulative ensquared energy fraction (dashed line, right axis). The PSF core is near-diffraction limited in the NIR, but a substantial fraction of the energy is contained in the extended halo. The PSF intensity is plotted in log scale.}\label{fig:psf}
\end{figure}

The AO PSFs were generated using detailed Monte Carlo AO performance simulations tailored to the E-ELT for both single-conjugate and laser tomographic AO (SCAO and LTAO, respectively), assuming an on-axis scientific target. The AO simulations provided the PSFs at a number of reference wavelengths, from which they are parameterised into base functions and thus extrapolated over the required wavelength range of the HARMONI simulation \citep{Zieleniewski2013}. In Figure~\ref{fig:psf} we show a slice through the H-band PSF together with the cumulative encircled energy (EE) fraction as a function of distance from the PSF centre, showing the rather complex shape of the simulated LTAO-corrected PSF. Whilst the core of the PSF is near-diffraction limited in the NIR, a substantial amount of energy is contained in the extended PSF halo. The PSF shown corresponds to a Strehl of aproximately 0.35.

Whilst the aim of HSIM is to provide ``realistic'' observations, its methodology is necessarily idealised in a number of ways. HSIM does not model the low level systematics associated with some NIR instruments, e.g. amplifier and row `bias', hot or bad pixels, sky line and continuum variation and related flexure (shift), persistence, or cross-talk. When pushing below the read out noise (RON), such systematics will likely become increasingly important. As~\citet{Zieleniewski2015} have shown, most modes of HARMONI will be RON-limited; as such, the simulations shown here assume that these systematics are either measured and corrected during reduction or sufficiently controlled in the instrument. Furthermore, it is assumed that the systematics are orders of magnitude below the RON of a single exposure, allowing one to `beat down' many factors below the RON with the addition of multiple exposures.

\section[]{Expected HARMONI Signal-to-Noise for Passive Galaxies at $z=2-4$}\label{sec:ryan}

Before simulating observations of one galaxy in detail, we perform a series of fiducial simulations to assess the feasibility and fidelity of \emph{integrated} absorption line spectroscopy of massive passive galaxies before the peak epoch of star formation. Details (and results) are given in Table \ref{tab:IntegratedSimProperties} and we briefly summarise here. 

For each of three redshifts (z=2, 3 \& 4) we simulate galaxies of three stellar masses ($\log M/$\msun = 10, 11 \& 12). Stellar populations are all assumed to have solar metallicity and ages are fixed at 3, 2 and 1 Gyrs for z = 2, 3 \& 4. In this way, the stellar populations are roughly as old as realistically possible, as the age of the Universe is 3.3 Gyrs, 2.1 Gyrs and 1.6 Gyrs respectively, at these epochs. We targeted the `traditional' Balmer-break region at rest wavelengths around 0.4~\micron~in all cases, as is common in high redshift studies of passive galaxies. Not only is this area rich in strong absorption lines for passive populations, such as Ca H\&K, but in younger populations the Balmer series becomes dominant and is a useful age indicator. 

At z = 2, 3 \& 4 the Balmer-break is well placed in the J, H and K bands allowing us to compare identical spectral regions in all cases. Standard exposure time calculators generally assume point source light profiles for simplicity; we test the validity of this assumption by modelling the galaxy light profiles as point sources, de Vaucouleurs profiles and Exponential profiles. The effective radii of the de Vaucouleurs and Exponential profiles are fixed at 0\farcs2 and 0\farcs5 respectively, for all redshifts. Although this leads to different physical sizes, the differences are small: 0\farcs2 equates to 1.7 kpc, 1.6 kpc and 1.4 kpc at z = 2, 3 \& 4; 0\farcs5 equates to 4.3 kpc, 3.9 kpc \& 3.6 kpc at the same redshifts. Furthermore, these sizes are typical of the early-type and late-type populations at high-z~\citep[see e.g.][]{vanderWel2014} and the large scatter in the mass-size relation more than covers for this small variation in physical size resulting from a fixed size on the sky. 

In the case of a point source, we know from previous simulations by~\citet{Zieleniewski2015} that the best S/N is achieved using the 20 $\times$ 20 mas spaxel scale. We note that the point source case is largely included here for comparison with the ETC calculation; whilst HARMONI will perform point source observations, almost all galaxies will be spatially resolved by the instrument. Unsurprisingly, we found the best S/N for extended sources was achieved with the largest spaxel scale of 30 $\times$ 60~mas. Individual exposures are simulated to be 900 seconds and we observe for a total of 10 hrs on-source. The integrated spectra were optimally extracted following~\citet{Horne1986} to optimize S/N.

Table \ref{tab:IntegratedSimProperties} lists our main results for integrated spectroscopy of passive galaxies beyond the peak epoch of star formation. Three S/N values are reported: that from an HSIM simulation assuming a point source; that from an HSIM simulation using a de Vaucouleurs light profile; and the S/N from an HSIM simulation using an Exponential light profile. Note that the results from the HSIM point source simulations were verified against a more traditional analytic exposure time calculation, and found to be consistent. Further verification of HSIM's output can be found in~\citet{Zieleniewski2015}.

\begin{table*}
   \centering
   \begin{tabular}{c|c|c|c|c|c|c|c} 
     \hline
	  Redshift    & Stellar Mass & Age & Magnitude & HSIM S/N & HSIM S/N & HSIM S/N \\
     (z)   &   ($\log M/$\msun) & (Gyrs) & (AB) & (PS) & (dV) & (Exp) \\
     \hline
	2 & 10 & 3 & J=26.85 & 3   & 1.4 & 0.9 \\
	3 & 10 & 2 & H=27.06 & 5   & 1.2 & 0.6 \\ 
	4 & 10 & 1 & K=26.27 & 3   & 0.6 & 0.4 \\ 
	\hline
	2 & 11 & 3 & J=24.35 & 26  & 15  & 9   \\
	3 & 11 & 2 & H=24.56 & 37  & 12  & 6   \\
	4 & 11 & 1 & K=23.77 & 30  & 6   & 3   \\
	\hline
	2 & 12 & 3 & J=21.85 & 141 & 125 & 85  \\
	3 & 12 & 2 & H=22.06 & 186 & 72  & 65  \\
	4 & 12 & 1 & K=21.27 & 195 & 47  & 26  \\
   \end{tabular}
   \caption{Summary of the simulated properties and S/N measures for passive galaxies of various masses, ages and redshifts, observed in 10 hrs of on-source exposure with HARMONI on the E-ELT. Reported HSIM S/N values are the averages between 0.4~$\mu$m and 0.45~$\mu$m (rest-frame). Point source (PS) calculations were performed using the 20$\times$20~mas spaxel scale, while de Vaucouleurs (dV) and Exponential (Exp) calculations were performed using the 30~mas$\times$60~mas spaxel scale; such spaxel scales are the optimal choice in the two circumstances. All models assume solar metallicity.}
   \label{tab:IntegratedSimProperties}
\end{table*}

\begin{figure*}
    \includegraphics[width=16cm]{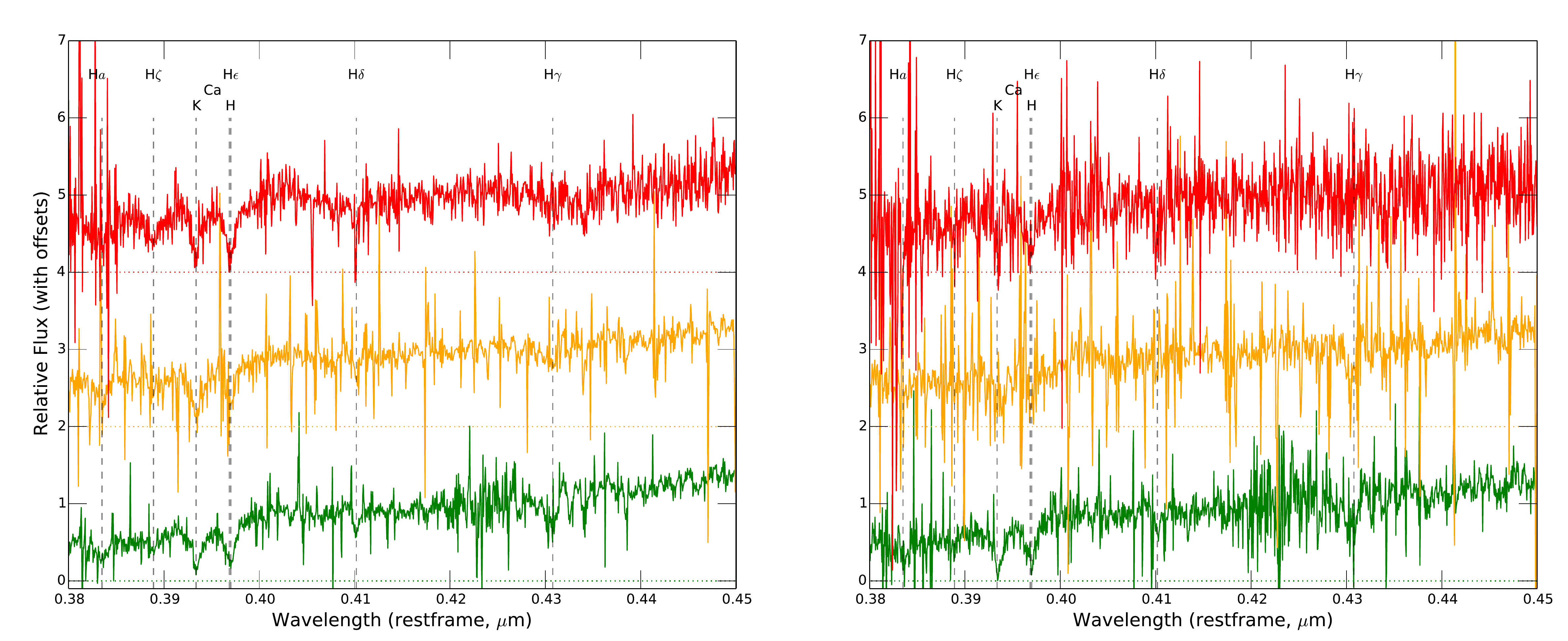}
    \caption{Simulated HARMONI spectra of passive galaxies for various redshifts (vertically offset) and light profiles (left and right panels). We show the optimally-extracted rest-frame spectra around the Ca H + K region for galaxies of stellar mass $\log M/M_{\sun}=11$ at z = 2, 3 and 4 (green, orange and red lines), observed in the J, H and K bands, respectively. Left: galaxies modelled with de Vaucouleurs light profiles. Right: galaxies modelled with exponential light profiles. Note that the S/N depends more on the shape of the light profile than the redshift, as confirmed in Table~\ref{tab:IntegratedSimProperties}: spectra on the left have visibly better S/N than on the right.}
    \label{fig:ryan}
\end{figure*}

The results shown in Table~\ref{tab:IntegratedSimProperties}, and plotted in Figure~\ref{fig:ryan}, allow us to make predictions for the feasibility of IFS observations of passive galaxies at high redshifts. The S/N numbers show that the S/N for realistic extended galaxy profiles are far lower than for the point source geometry. Using realistic light profiles, the data suggest that galaxies with 10$^{10}$~\msun~stellar mass will remain out of reach with HARMONI in a 10-hour on-source exposure. At the highest masses, the S/N values are likely to be just high enough for \emph{resolved} studies. These galaxies are at the extreme upper end of the known mass function at high redshifts, they therefore do not represent the ``typical'' population~\citep{Marchesini2009, Marchesini2010}. 

These conclusions come with several caveats.~\citet{Mendel2015} recently showed that the SSP-equivalent stellar age of passive galaxies plateaus around 1 Gyr beyond $z=2$. The populations we assumed for these simulations of passive galaxies are therefore likely too old, and our results too pessimistic: younger populations are more luminous and will therefore give a higher S/N for a given exposure time. On the other hand, the simulations do not model the low-level systematics that can be present in NIR systems; such systematics may limit the sensitivity of the instrument and reduce the achievable S/N.

\section[]{Detailed mock HARMONI observations of a $z=3$ star-forming galaxy: Input data and Methodology}\label{sec:siminput}

Our initial investigation into HARMONI's performance for IFU studies of high-z galaxies made a number of simplifying assumptions, such analytically described galaxy morphologies, simple star formation histories and solar metallicities. Present-day cosmological simulations provide us with the opportunity to generate far more realistic model galaxies under fully-characterised physical conditions. In the following sections, we describe a first investigation into HARMONI's performance in studying the stellar kinematics at $z=3$ based on a galaxy extracted from a cosmological simulation. This study is a ``pathfinder'' study to develop a pipeline for mock IFU observations of simulated galaxies with HARMONI.

\subsection{The \ramses \nutfb~simulation}\label{sec:nutfb}

Present-day cosmological simulations can now reach sufficiently high spatial resolutions to make useful observational predictions. In the following sections, we present a proof-of-concept study into the feasibility of observing galaxies at high redshift with HARMONI, based on the properties of a galaxy from a cosmological simulation. The starting point for this study is a cosmological simulation performed with the Eulerian AMR code~\ramses\citep{Teyssier2002}. 

The data were extracted from the \nut simulations, described in detail by~\citet{Powell2011, Kimm2011}. \nut is a suite of high resolution cosmological simulations of a Milky Way-like galaxy forming at the intersection of three filaments in a $\Lambda$CDM cosmology. The values of the input cosmological parameters are consistent with a WMAP5 cosmology~\citep{Dunkley2009}, see Table~\ref{tab:cosmo}. Reionization of the Universe is simulated by switching on a uniform UV radiation field at high redshift ($z_{UV}$), after which atomic cooling causes gas to cool to 10$^4$~K~\citep{Sutherland1993}. When the density in a gas cell exceeds a threshold density $n_{th}$, star particles are spawned by a Poisson process following a standard Schmidt law. A star formation efficiency of 1\% per free-fall time is assumed. The mass of the star particles is determined by a combination of the minimum grid size and the value of $n_{\rm th}$.

For this work we use the output from one of the three resimulations performed at higher spatial resolution than the full suite, propagated to lower redshift. \nutfb~terminates at $z=3$ and reaches a physical spatial resolution of 12 pc. At this resolution the mass of each dark matter particle is 5.5 $\times$ 10$^4$ M$_{\sun}$, and each star particle represents a stellar population of mass $\sim$2 $\times$ 10$^4$ M$_{\sun}$ at birth. 

In addition, \nutfb~includes the effect of supernova feedback. After 10 Myr, massive stars are assumed to undergo Type II supernova explosions, releasing 50\% of their energy (10$^{51}$ ergs) into the interstellar medium (ISM) as thermal energy, and 50\% as kinetic energy. As a result, the star particles undergo mass loss of 10.6\%, assuming a Salpeter initial mass function (IMF;~\citealt{Salpeter1955}).

Every supernova bubble with an initial radius of 32 pc sweeps up the same amount of gas as that initially locked in the star particles. The supernova feedback prescription is described in more detail by~\citet{Dubois2008}. The values for all input parameters for the \nutfb simulation are shown in Table~\ref{tab:input}. Simulation output was processed using the Python package pynbody~\citep{pynbody}.

\begin{table}
    \centering
    \begin{tabular}{|l|c|}
        \hline
        \multicolumn{2}{|c|}{NUT input cosmology}\\
        \hline
        $\Omega_M$ & 0.258\\
        $\Omega_{\Lambda}$ & 0.742 \\
        $H_0$ & 72 km s$^{-1}$ Mpc$^{-1}$ \\
        \hline
    \end{tabular}
    \caption{Input cosmology for the NUT and NutFB simulations. $\Omega_M$ and $\Omega_{\Lambda}$ are the matter and vacuum energy density parameters, $H_0$ is the Hubble constant. }\label{tab:cosmo}
\end{table}

\begin{table}
    \centering
    \begin{tabular}{|l|c|}
        \hline
        \multicolumn{2}{|l|}{NutFB simulation parameters and physical ingredients}\\
        \hline
        L (Mpc/h) & 9 \\
        $\Delta x_{\rm min}$ (pc) & 12 \\
        $m_{\rm DM}$ (M$_{\sun}$) & 5 $\times$ 10$^4$ \\
        $m_{\rm star}$ (M$_{\sun}$) & 2 $\times$ 10$^4$ \\
        $n_{th}$ (H cm$^{-3}$) & 400  \\
        $z_{\rm end}$ & 3 \\
        $z_{\rm UV}$ & 8.5 \\
		$\eta$ & 1\% \\
        \hline
    \end{tabular}
    \caption{Summary of parameters and physical ingredients for the~\nutfb~simulation from which our data are extracted~\citep{Kimm2011}. L is the total simulated (co-moving) volume; $\Delta x_{\rm min}$ the physical spatial resolution; $m_{\rm DM}$ and $m_{\rm star}$ the mass of dark matter and star particles, respectively; $n_{th}$ the volume threshold for star formation; $z_{\rm UV}$ and $z_{\rm end}$ the redshifts at which reionization is introduced and at which the simulation is terminated. $\eta$ is the star formation efficiency.}\label{tab:input}
\end{table}

\subsection{A simulated star-forming galaxy at $z = 3$}

The~\nutfb~simulation focuses on a galaxy located at the centre of a 10$^{11}$ M$_{\sun}$, $\sim$35 kpc halo, which displays signs of disk formation and is seen at an inclination of approximately 45\degree. The galaxy's nearest companions are found at distances of 7 and 10.5 kpc; these satellites are not interacting with the main galaxy and their masses are small relatively to its mass. The gas, star and dark matter particles of the galaxy were extracted from the simulation volume in a 5 kpc radius spherical volume, representing $\sim$2 times the half-mass radius.

\begin{figure*}
    \centering
    \includegraphics[width=\textwidth]{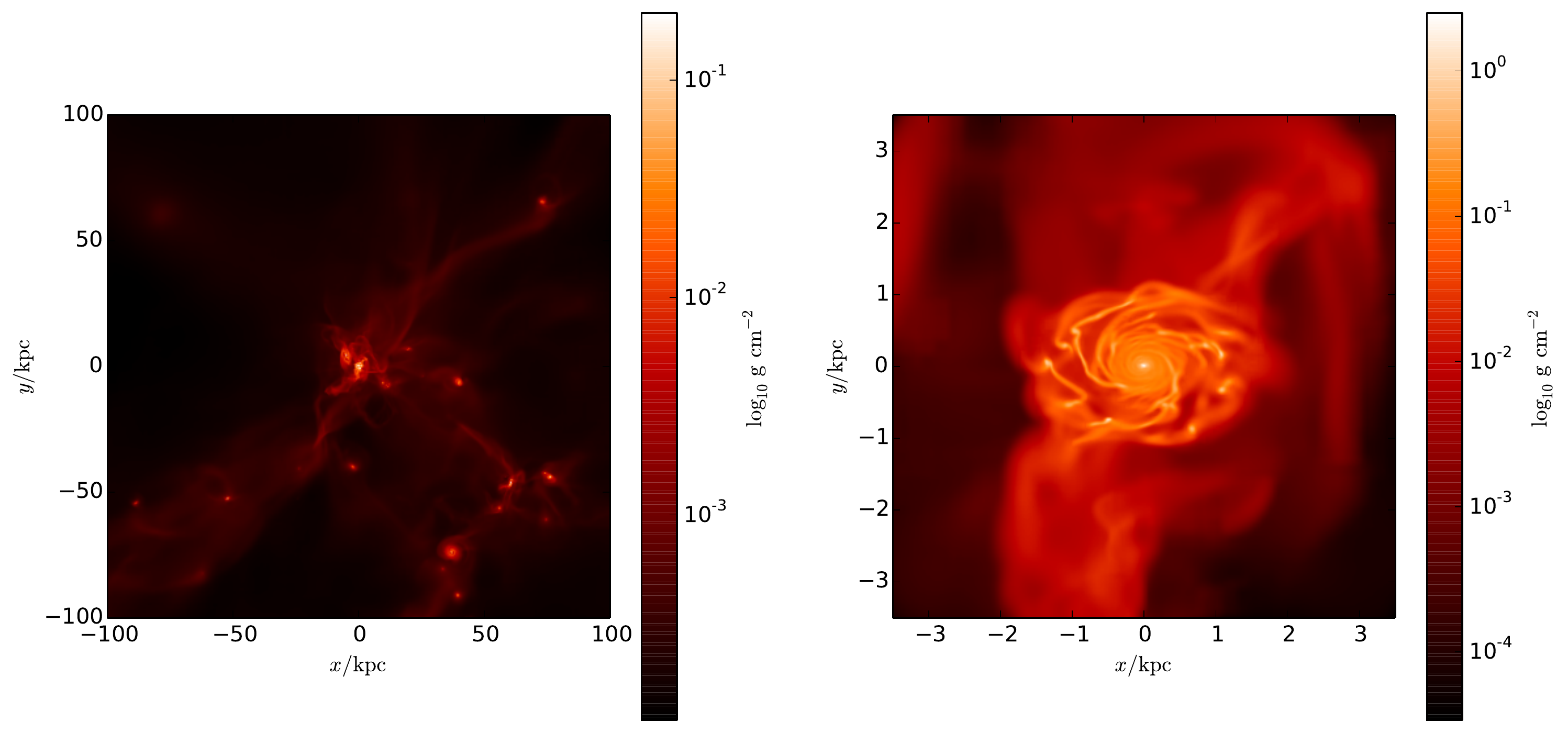}
    \caption{Left: Dark matter density plot of the 10$^{11}$~M$_{\sun}$ halo and its surroundings. The image covers 100 $\times$ 100 kpc, the halo itself measures approx. 35 kpc in radius. Right: Zoom-in on the galaxy at the halo's centre. The gas surface density shows a clear spiral structure with a number of star-forming knots associate with the spiral arms.} \label{fig:overview}
\end{figure*}

\begin{figure}
    \centering
    \includegraphics[width=8cm]{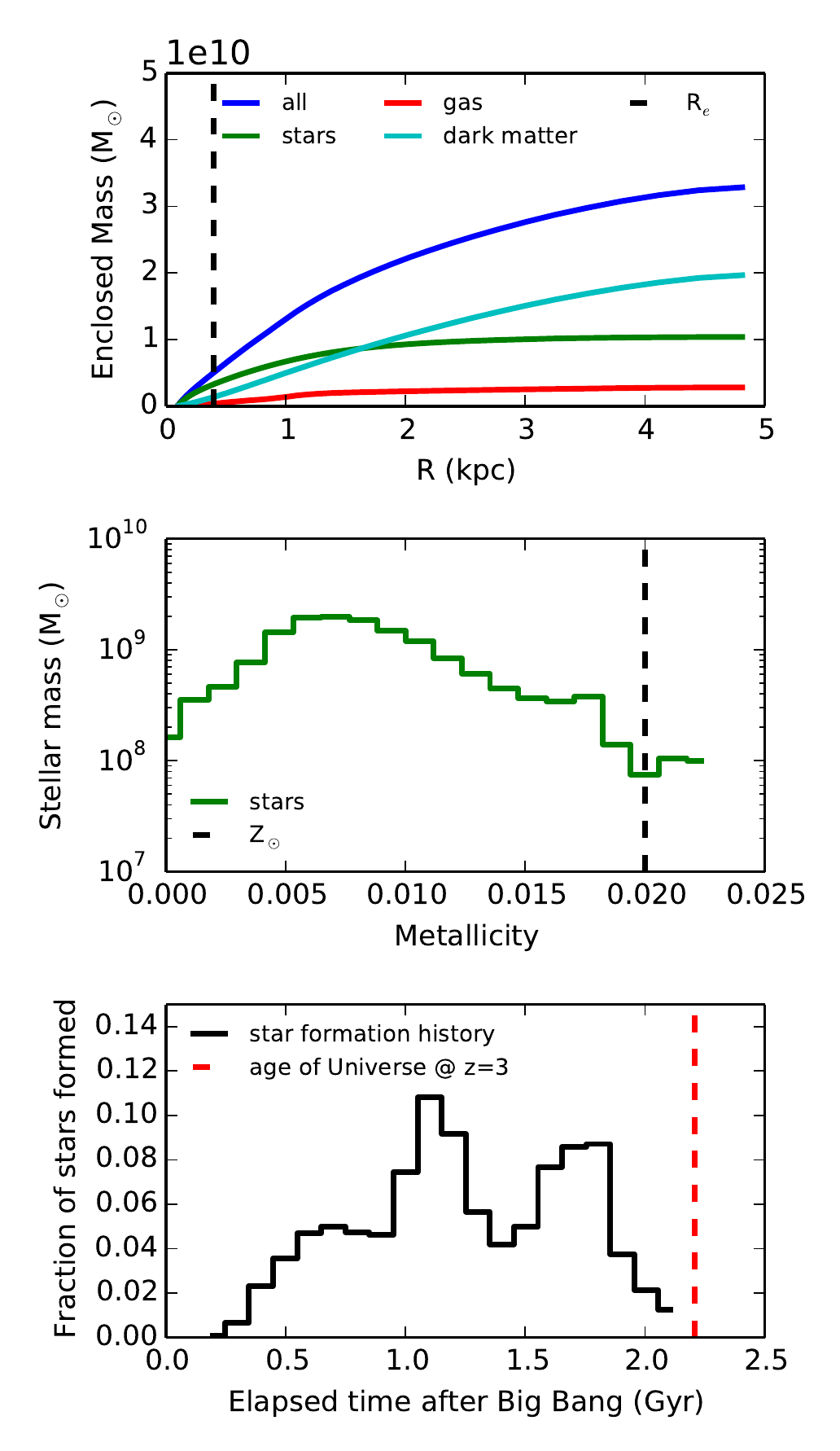}
    \caption{Overview of the~\nutfb~$z=3$ galaxy. Top: the cumulative radial mass profile, for gas, stars, dark matter as well as the sum of all 3 constituents, for the 5 kpc radius extracted volume. Middle: the distribution of star particle metallicities; the dashed line indicates the solar metallicity. Bottom: the galaxy's star formation history; the age of the Universe at $z=3$ (2.2 Gyr) is shown with the dashed line, this represents the maximum stellar age at that redshift.}\label{fig:galaxy_profiles}
\end{figure}

Figure~\ref{fig:overview} shows an image of the galaxy in dark matter and gas column density, and in Figure~\ref{fig:galaxy_profiles} we show a number of plots describing the physical state of the galaxy: the enclosed mass profiles for the full galaxy and its baryonic and dark matter constituents; its metallicity distribution; and star formation history. The total stellar mass of the galaxy is $\sim$1 $\times$ 10$^{10}$ M$_{\sun}$, contained in some 7 $\times$ 10$^5$ particles. We note that this stellar mass is approximately an order of magnitude higher than expected from the typical stellar mass-halo mass relations for a 10$^{11}$ M$_{\sun}$ halo at z = 3~\citep{Behroozi2010, Behroozi2013}. This discrepancy is due to the relative inefficiency of the feedback prescription in the~\nutfb~simulation, which has allowed stars to form more rapidly than is known to have occurred in real systems. Another consequence is that the bulk of the stellar mass formed at unrealistically low metallicity (see Figure~\ref{fig:galaxy_profiles}). \citet{Ilbert2013} find a characteristic mass $\log{M_*}$ at $z=3-4$ of $10.74^{+0.44}_{-0.20}$~\msun, placing the \nutfb~galaxy around 0.2 $M_*$ (with a range of 0.07 - 0.3 $M_*$) on the galaxy stellar mass function around $z=3$.

Whilst its mass and star formation history make the galaxy not optimally representative of the known population at this redshift, it was chosen for these simulations based on the high spatial resolution of the \nutfb~simulation, which is important for our simulations of spatially resolved spectroscopy with HARMONI. We note also that the \nutfb~simulation is focused on this particular halo and terminates at $z=3$; choosing a more massive galaxy, or simulating the galaxy at lower redshift was therefore not possible. In Section~\ref{sec:obscontext} we provide further discussion of the galaxy in observational context.

In the following sections we describe the method used for converting the~\nutfb~simulation data into a spectral cube of the galaxy's stellar population. Using stellar population synthesis models, we first produced a rest-frame visible spectrum, integrated over the entire stellar population. Based on this, we identify the best wavelength regions to target with HARMONI observations and produced input spectral cubes covering these regions.

\subsection{Converting star particles into starlight}

The RAMSES particle output forms the starting point for modelling the light from the galaxy, with each star particle representing a simple stellar population (SSP) characterised by a mass, age and metallicity. In this paper we focus on the light from stars only to study the stellar kinematics in the galaxy. 

We use the MIUSCAT stellar population synthesis models~\citep{Vazdekis2010, Vazdekis2012}, assuming a Salpeter-like bimodal initial mass function (IMF) with a slope of -1.30, to create the galaxy's stellar light spectrum. Below 0.6~\msun~the IMF slope flattens following a recipe described in~\citet{Vazdekis1996}. The MIUSCAT model spectra cover a wavelength range from $\sim$3500 to $\sim$9500\AA. Each star particle is matched up with a model SSP according to its age and metallicity, scaled to the particle mass, and adjusted for its line-of-sight velocity. At optical wavelengths these spectra are sampled at 0.9\AA~with a spectral resolution (FWHM) of 2.5\AA~\citep{Beifiori2011}. 

We note the discrepancy between assumed IMF between the MIUSCAT library (the bimodal IMF) and the~\nutfb~simulation, where the (unimodal) IMF determines the star particles' mass loss from supernovae at an age of 10 Myr. For a given mass, the latter will contain a larger fraction of low-mass stars than a bimodal-IMF population. In the MIUSCAT stellar spectra with bimodal IMF the OB stars will therefore occupy a higher fraction and the SSP will undergo a greater mass loss than the 10.6\%. This is not expected to change the conclusions of this study.

The choice of stellar library may affect the spectral shape and the galaxy's luminosity, particularly in the context of the contribution of thermally pulsating asymptotic giant branch (TP-AGB) stars~\citep{Maraston2005, Tonini2009}. This stellar evolution phase can account for a significant fraction of the luminosity of stellar populations of 0.3 to 2 Gyr in age. Given the age of the Universe at $z=3$ (2.2 Gyr) and the wavelength coverage of HARMONI, an accurate inclusion of the effects of TP-AGB stars is likely to affect the galaxy's spectrum redwards of 5000~\AA and particularly in the rest-frame NIR. As the simulations presented here focus on the rest-frame visible wavelength range, this will not affect our results.

For our initial investigation into the observational strategy with HARMONI, an integrated rest-frame spectrum was produced for the galaxy by summing the spectra of all star particles in the extracted volume, and redshifted to $z=3$. As shown in Figure~\ref{fig:galaxy_profiles}, the galaxy is actively star-forming, the maximum stellar age is just 2 Gyr, and the galaxy has a substantial stellar population with ages less than 1 Gyr; while we do not model the galaxy's nebular emission, strong emission is expected in the hydrogen lines, filling in the absorption features. This will complicate the analysis of spectral absorption lines in a real system. 

\subsection{A $z=3$ galaxy integrated spectrum}

\begin{figure*}
    \centering
    \includegraphics[width=\textwidth]{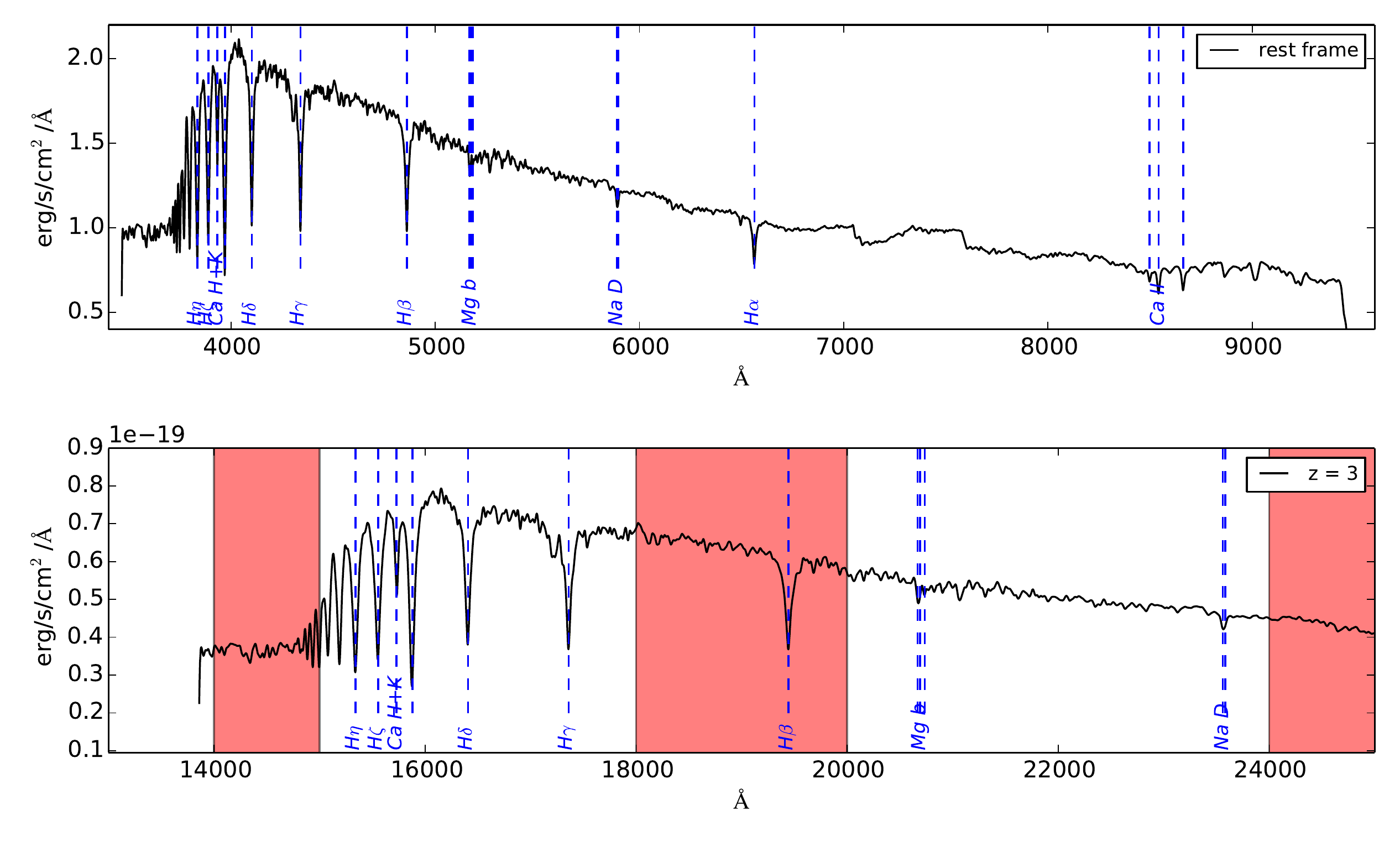}
    \caption{Integrated stellar spectra of the $z=3$ \ramses~galaxy based on the physical properties and line of sight velocities of $\sim$7 $\times$ 10$^5$ star particles. Top: the rest-frame visible wavelength spectrum from $\sim$3500 to $\sim$9400\AA, with key absorption lines indicated in the dashed lines. Bottom: The same spectrum redshifted to $z=3$. The inter-band regions are shaded in red; the visible spectrum is redshifted into the H and K bands, and beyond into the thermal IR. The same spectral lines as in the rest frame spectrum are marked and labelled here.}\label{fig:intspec_plots}
\end{figure*}

The integrated spectrum of the galaxy's starlight is shown in Figure~\ref{fig:intspec_plots}, both in  the optical rest frame and shifted to $z=3$. From these spectra we calculate a rest-frame V-band magnitude of -21.4. At $2.7\leq z < 3.3$ ~\citet{Marchesini2012} find a characteristic rest-frame V magnitude ($M_V^*$) of -22.83$^{+0.18}_{-0.21}$ from the best-fit galaxy luminosity function, placing our galaxy at $\sim$0.3$L_*$. 

The hydrogen Balmer lines are prominently seen in absorption, however, in a young star-forming system these lines may have substantial overlying emission from star-forming regions (e.g.~\citealt{Steidel2014}). For $2.7 < z < 3.1$ the strong H$\beta$ line lies between the H and K band windows, and can therefore not be used for analysis of either passive or active galaxies near these redshifts using ground-based telescopes. At $z=3$ the K band, which has good transmission and relatively few sky and telluric lines, contains the Mg I b triplet ($\lambda \lambda$ 5167, 5173, 5184 \AA~rest frame) and the Na D doublet ($\lambda \lambda$5890, 5896 \AA), which are relatively weak in our spectrum. 

The NIR H band, covering 1.5-1.8~\micron (3750-4500 \AA~rest frame), contains the higher-order Balmer lines (H$\gamma$-$\eta$), as well as the Ca II H+K lines. The Balmer lines are useful for dispersion and stellar population analyses for quiescent systems, but not for star-forming galaxies such as our \ramses galaxy. The Ca Triplet (CaT hereafter) around 8500\AA, typically the ``gold standard'' in the optical bandpass for these measurements~\citep{Greene2006}, is shifted into the thermal IR L-band, inaccessible to HARMONI and challenging from the ground.

The only features available in our spectrum for stellar velocity dispersion ($\sigma_{*}$) measurements are the Mg I b lines in the K band and the Ca H+K lines in H. Use of these lines for dispersion measurements in the presence of nebular emission is problematic, as extensively discussed in the literature~\citep{Kormendy1982, Greene2006, Blanc2013}. ~\citet{Blanc2013} model the use of the Mg I b line for measuring $\sigma_{*}$ for simulated spectra for different types of stellar populations, S/N levels and intrinsic dispersions. They find that the most accurate $\sigma_{*}$ can be measured from these features if fitted over a narrow wavelength range surrounding the lines, with systematic offsets seen in low S/N and low $\sigma_{*}$ observations and at the cost of a larger random error. The systematics are stronger in star-forming or post-starburst stellar populations. 

Similarly, \citet{Greene2006} give an in-depth study of $\sigma_{*}$ measurements using the Ca H+K lines. The lines are located near the 4000 \AA~Balmer break, where the continuum gradient is steep. The gradient and the H+K line shapes are strongly dependent on spectral type and therefore sensitive to template mismatch. In addition, as Ca H overlaps with the H$\epsilon$ line, it is nearly always observed shifted away from its expected rest wavelength and unreliable for fitting. This leaves just the Ca K line. Finally, as the lines are intrinsically broad, the $\sigma_{*}$ shows systematic offsets when compared with the CaT measurement, particularly in low $\sigma_{*}$ systems.

Based on these considerations, we will focus on the spectral windows around the Ca H+K doublet (3800-4300 \AA) and the Mg I b triplet (5000-5500 \AA) for the mock observations with HARMONI and the subsequent analysis. Separate spectral cubes were created for each range. The benefit of using a galaxy extracted from cosmological simulations is that we have full knowledge of the input physics to verify the results from the analysis. In the following section we describe how the spectral cube was generated for the galaxy, which we use as input for the instrument simulator.

\subsection{The model spectral cube}

The principle for generating a spectral cube from the star particle data follows the recipe described for the integrated spectrum, with the spatial information added. Each particle is described by a physical (x,y) position, assuming the z-axis represents the line of sight. The simulation was centred on the halo's centre of mass, and positions were converted to use this coordinate as zero-point. The distances were then converted to angular positions on sky using the co-moving distance at that redshift for the cosmological parameters of the simulation. 

To allow the instrument simulator to convolve the input cube to the 10 $\times$ 10 mas spaxel scale of HARMONI, the input cube was sampled 10 times more finely, on a 1 $\times$ 1 mas grid of pseudo-spaxels. We co-added the (rest-frame) spectra of the star particles found in each pseudo-spaxel using the MIUSCAT models as described above~\citep{Vazdekis2010}, matching their ages and metallicities with the library spectra, scaling for their mass, and corrected for their individual line of sight (z-axis) velocities. All spectra were interpolated onto a common wavelength grid, and finally redshifted to $z=3$. The resulting spectral sampling is 3.6\AA~and the FWHM 10~\AA~in the observed frame, giving R$\sim$1500-2500 across the H and K bands. We note that this is lower than the spectral resolving power for a typical observation with HARMONI (R$\sim$3500); the spectral resolution of the output data will in this case be limited by the input data.

The flux was scaled to account for the distance and redshift dimming using the following:

\begin{equation}
	F_{\rm \lambda,z}  = F_{\rm \lambda,0} \times \left(\frac{10 \textrm{pc}}{d_L}\right)^2 \left(\frac{1}{1+z}\right) \times 10^6
\end{equation}

\noindent where $F_{\rm \lambda,z}$, $F_{\rm \lambda,0}$ are the redshifted and rest frame fluxes, respectively; $d_L$ is the luminosity distance. The factor 10$^6$ converts the flux in 1 mas from erg s$^{-1}$ cm$^{-2}$ \AA$^{-1}$ to erg s$^{-1}$ cm$^{-2}$ \AA$^{-1}$ arcsec$^{-2}$, the flux density required by HSIM. The cosmological parameters give a spatial scale of 0.1278\arcsec/kpc at $z=3$; our 10 kpc sphere therefore occupies a $\sim$1.3\arcsec~field. 

To limit the processing times, we restrict the wavelength range around the features of interest, as described in the previous section. The ranges selected are 3800-4300 \AA, which is redshifted into the H-band, and 5000-5500 \AA, redshifted into the K-band; we thus have two individual spectral cubes for the galaxy that will be used as HSIM input. At R$\sim$3500, we note that HARMONI will be able to cover the full H and K bands simultaneously, providing extremely rich data in a single exposure. 

\begin{figure*}
    \includegraphics[width=13 cm]{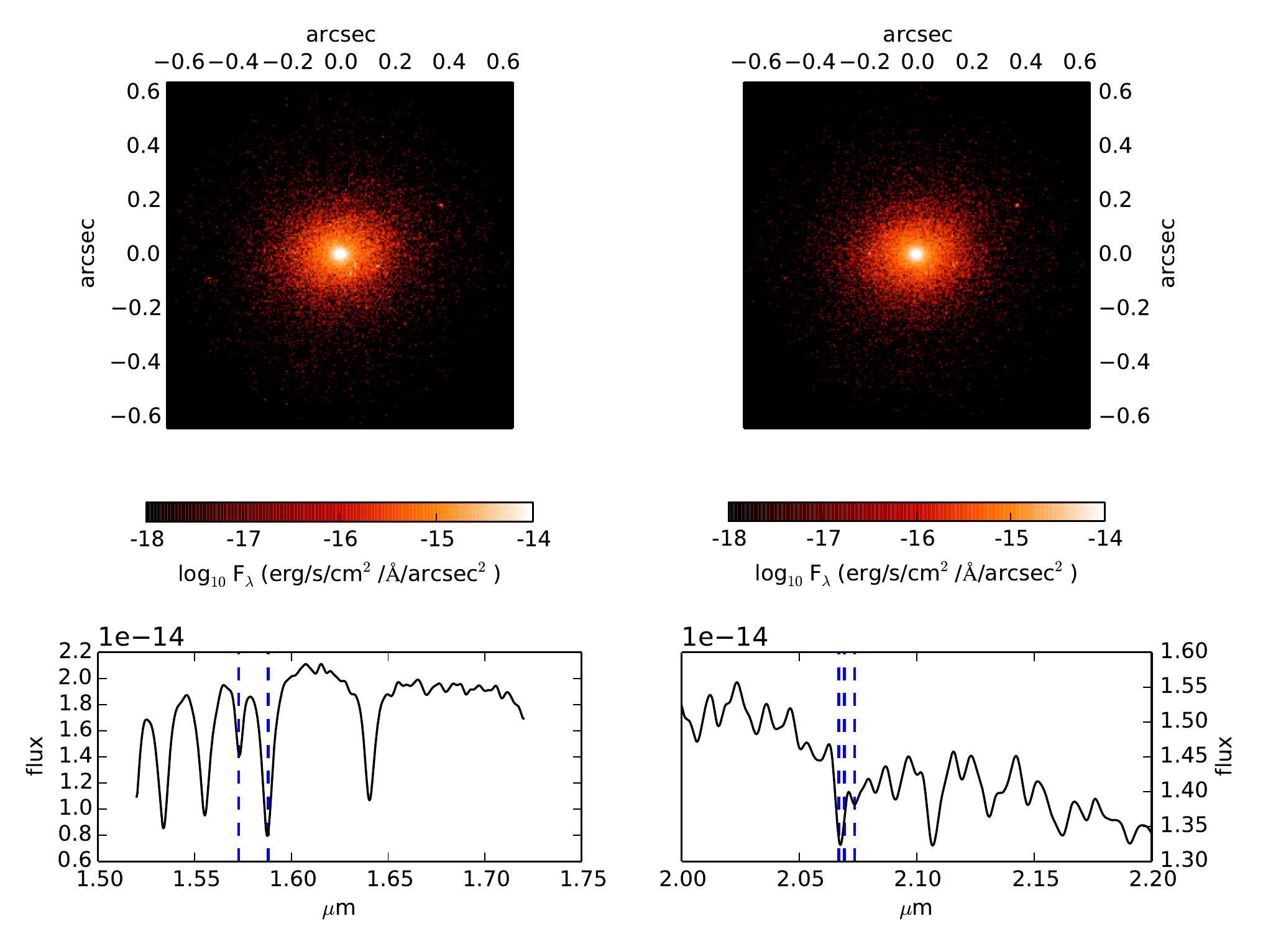}
    \caption{Views of the two face-on input cubes created for the \ramses galaxy. The top row shows integrated images of the full wavelength ranges of both cubes, covering the Ca H+K region (1.52-1.72~\micron; left) and Mg I b region (2-2.2~\micron; right). The image is log-scaled in flux, in units of erg s$^{-1}$ cm$^{-2}$ \AA$^{-1}$ arcsec$^{-2}$, and each pixel represents 1 $\times$ 1 mas on sky. The bottom row shows spectra from each cube extracted and summed over 20 $\times$ 20 spaxels (20 $\times$ 20 mas or $\sim$0.15 $\times$ 0.15 kpc). The spectral features of interest are marked in eah spectrum with the dashed lines: the Ca H+K lines in the left-hand panel, the Mg I b triplet on the right.}
\label{fig:inpcube_grid}
\end{figure*}

In Figure~\ref{fig:inpcube_grid} we show images of the Ca H+K (H band) and Mg I b (K band) cubes, integrated over their full respective wavelength ranges (1.52-1.72~\micron~and 2.0-2.2~\micron), and each accompanied by a spectrum extracted from their central 20 $\times$ 20 spatial pixels. No smoothing has been applied to this image, star particles are collected as point sources in each 1 mas pseudo-spaxel. The instrument simulator will convolve the image with appropriate detailed PSFs. 

In addition to the 1 mas-sampled cube, a second cube was produced using the exact same method as described above, sampled on a 10 $\times$ 10 mas pseudo-spaxel grid to match the HARMONI spatial scale. This cube was produced for the sole purpose of providing a benchmark for the data analysis of the simulated cubes.

\section{Simulated observations with HARMONI}\label{sec:hsims}

The spectral cubes constructed for the $z=3$ \ramses galaxy in 2 key NIR wavelength ranges, targeting the Ca H+K and Mg I b absorption lines, were subsequently passed to HSIM to produce mock observed output cubes. The observational parameters are summarised in Table~\ref{tab:siminput}. Assuming median seeing conditions, a moderate zenith angle of 30\degr, and an LTAO PSF, we observe the galaxy with 60 exposures of 900s each, totalling 15 hours of on-source time. We use the 10 $\times$ 10 mas spaxel scale and observe the full field of view of the cube, of 1.3\arcsec.

The spatial scale should be chosen on a case-by-case basis depending on the target and the required measurement, and~\citet{Zieleniewski2015} show the sensitivity predictions for different wavebands and spatial scales with HARMONI. For point sources or extended sources for which only integrated spectroscopy is needed, the 20 $\times$ 20 mas scale provides better sensitivity than the 10 $\times$ 10 mas. In this case however the goal is to investigate HARMONI's performance for \emph{spatially resolved} spectroscopy of the $z=3$ galaxy, and for this we simulate the observations at the finer 10 $\times$ 10 mas scale.

For this particular observation we assume a simple ABBA dithering scheme without needing off-source pointings for sky subtraction. For brighter or more extended targets additional pointings may be required, giving larger observational overheads for a given on-source exposure time. Recent work by~\citet{Thatte2012} demonstrates how to improve the quality of on-source sky subtraction techniques for IFS, improving the overall efficiency by factors of 2 to 4.

The nominal HARMONI spectral resolving power to be used for these observations is the R$\sim$3500 mode, which covers the full H and K band in one exposure. However, as the MIUSCAT model spectra have a lower intrinsic resolution (R$\sim$1500-2500), we allow the software to maintain the resolution of the input spectral cube and the spectrum is not affected by the instrumental line broadening. In the spatial dimensions, the software convolves the input cube with realistic AO-corrected PSFs, which are generated for each wavelength in the cube. Whilst HSIM has the ability to include the effects of ADR, we do not consider this effect in our simulations.

HSIM is designed to mimic as closely as possible a real observation, including realistic noise, background and transmission properties, so as to enable ``realistic'' data processing and analysis on the output data. To this end, it returns a sky background cube, and a transmission cube as may be obtained from a standard star in an observing programme. To aid with the comparison of the mock observations to the input data and investigate the effect of PSF convolution on the recovered kinematics in particular, a second simulation run was performed in which the PSF convolution was switched off. Analysis of the integrated galaxy spectrum allows us to compare with the initial simulations presented in Section~\ref{sec:ryan}. Finally, after finding the S/N of the observed spectral cube too low to derive the spatially resolved kinematics of the galaxy, we simulated a 15-hour observation (on-source time) of the galaxy with its flux boosted by a factor of 100, respresenting the high S/N limit for HARMONI.

For all simulations we obtain from HSIM the ``true'' observed cube as well as a noiseless data cube, which represents the counts from the object in the absence of photon noise or sky background. The noiseless output allows to us to compare the PSF and no-PSF simulations against the input data, and demonstrate the effect of the convolution on the recovered kinematics.

\begin{figure*}
    \includegraphics[width=13cm]{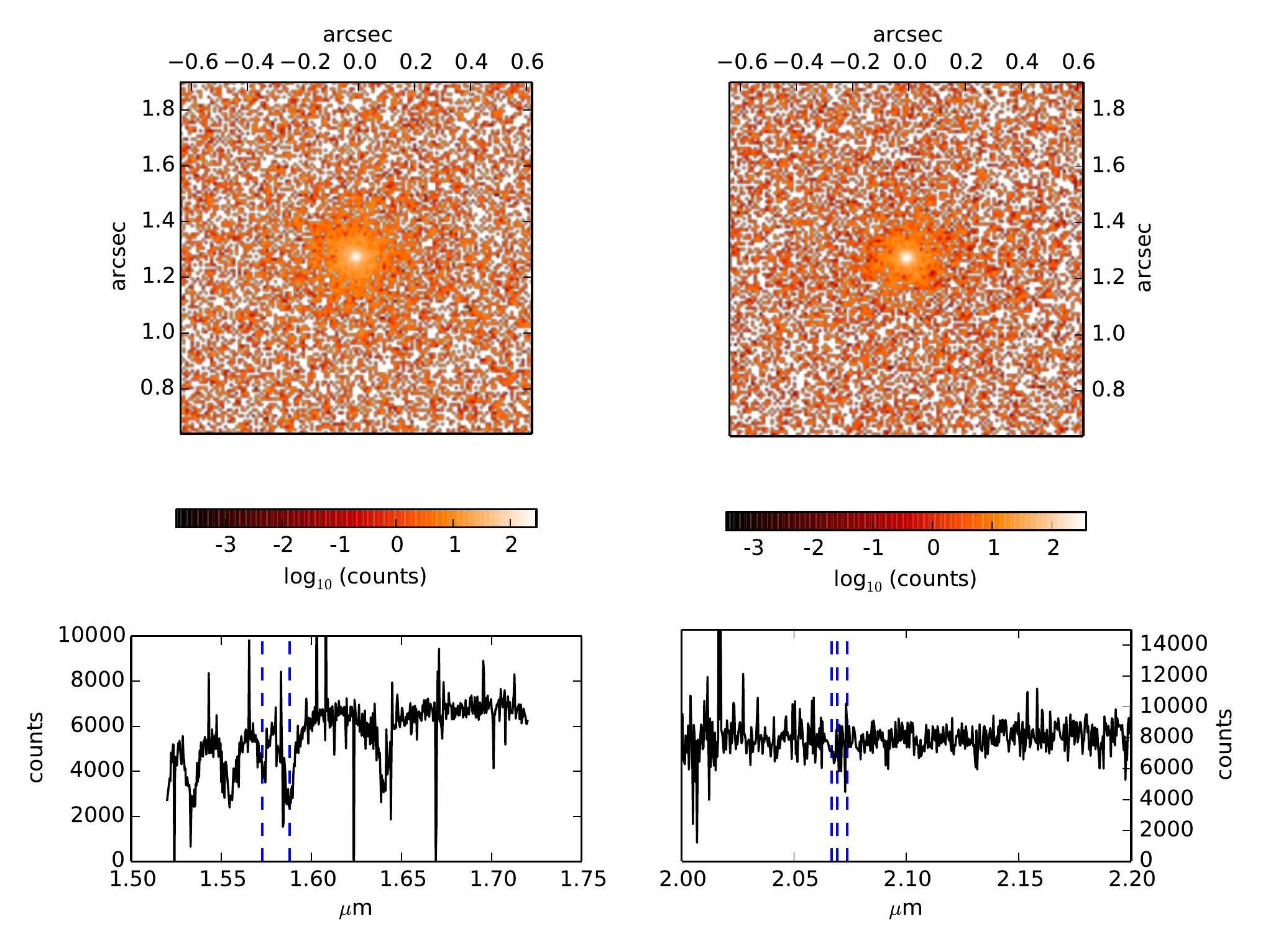}
    \caption{Overview of the simulated HARMONI observations of the $z=3$ \ramses galaxy (including PSF convolution), showing equivalent views to those in Figure~\ref{fig:inpcube_grid}. Top panels: Integrated images of the observed spectral cubes covering the Ca H + K (L; 1.52-1.72~\micron) and Mg I b (R; 2.0-2.2~\micron) spectral regions; the cubes are background subtracted and divided by a transmission spectrum. Bottom panels: Spectra extracted from the central 2 $\times$ 2 spatial pixels of the output cube, corresponding to 20 $\times$ 20 mas on sky.}\label{fig:outcube_grid}
    \end{figure*}

Figure~\ref{fig:outcube_grid} shows integrated images of the output cubes \emph{with} PSF convolution and realistic noise, side by side for both wavelength ranges, accompanied by the extracted spectra from the central 2 $\times$ 2 spaxels; this is the equivalent to Figure~\ref{fig:inpcube_grid} for the simulator output cubes. The output cubes were background-subtracted and divided by a transmission spectrum before the median 2D image was produced. Note that differences between these spectra and those shown in Figure~\ref{fig:ryan} are largely due to the different assumptions used to produce the input data. For the passive galaxies study of Section~\ref{sec:ryan} we assume a single-burst star formation history and solar metallicity for the entire stellar population, i.e. the entire galaxy is represented y a single SSP spectrum; for the~\nutfb~galaxy each star particle is matched to an SSP according to its age, metallicity and intrinsic line-of-sight velocity. The spectra shown were also extracted using different methods.

\begin{table}
    \begin{tabular}{|l|c|}
        \hline
        HARMONI Simulator input parameters & \\
        \hline
        DIT (s) & 900   \\
        NDIT    & 60    \\
        Spaxel scale (mas)  & 10 $\times$ 10\\
        Seeing  & 0.67  \\
        Zenith distance (deg) &   30 \\
        AO Mode & LTAO \\
        \hline
    \end{tabular}
    \caption{Summary of input parameters for the simulated HARMONI observations of the $z=3$ \ramses galaxy.}\label{tab:siminput}
\end{table}

\section{Data analysis methods}\label{sec:data_analysis}

The two sets of output data cubes from the HARMONI simulator, covering the Ca H+K and the Mg b lines respectively, were used to analyse and study both the integrated spectrum of the galaxy as well as the spatially resolved spectral cube. All data processing and analysis was performed using common techniques and software tools, to mimic real observations as closely as possible. For the initial data processing, we used the simulator's output auxiliary files to perform realistic data reduction steps of sky subtraction and division by a transmission spectrum. The noiseless output cubes require only the latter step as they do not include the sky background. All spectral extractions performed in this portion of the study over small or large apertures were uniformly weighted. We note that different weighting schemes are possible when extracting or co-adding IFS data, each with their own resulting biases; we refer to~\citet{Davies2011} for a detailed discussion of this.

Whilst the strength of an integral field spectrograph is in the spatially resolved study of galaxies, such instruments are also well suited to integrated spectroscopy. With integral field spectroscopy, the spectrum is not subject to slit losses, and spectra can be binned spatially to improve SNR via simple co-adding or optimal extraction techniques.

To eliminate noisy background spaxels, a region measuring approximately 1.5 $\times$ the observed half-light radius (see Section~\ref{sec:noiseless_results}) was extracted for further analysis; this corresponds to 1/3rd of the full field, or 0.43\arcsec. An integrated spectrum was created for both noisy and noiseless cases by coadding the signal in all spaxels of the 0.43\arcsec~sub-region. We note that a well-designed optimal extraction procedure can boost the S/N by $\ge$10\% over standard extraction~\citep{Naylor1998}, however different extraction methods will introduce different biases into the analysis, which must be taken into consideration~\citep{Davies2011}. The final integrated spectrum was produced by stitching the two spectra together onto a common wavelength axis. The wavelength region for which no spectrum is available (to save on computational time) was set to zero and masked from further analysis.

From the integrated spectrum the stellar kinematics were determined using the penalised pixel fitting method of~\citet[][hereafter pPXF]{Cappellari2004}\footnote{The pPXF software is available at~\citet{ppxf_software}}, commonly used for kinematics analysis of both integrated and spatially resolved spectra, using the same MIUSCAT spectral templates that were used to generate the input spectral data. We restricted the templates used for the fitting to those with ages below the age of the Universe at redshift 3 (2.2 Gyr). The pPXF fitting routine allows the user to mask out regions of the spectrum, we use this facility to mask out the region between the two modelled cubes so this does not affect the fit. As the galaxy is star-forming, we exclude the H absorption lines from the fit, as these are likely to be seen in emission in the true system. For the integrated spectra we compare the effect of this masking by performing the fit with the H lines included, as may be seen in a passive galaxy at this redshift.

With the most restrictive line masking, we effectively have only the Ca K line in the H band and the Mg b lines in the K-band, the latter being very weak for our young and low-metallicity stellar population. The advantage of the pPXF method over simple spectral line profile fitting is the full-spectrum fitting approach, which uses both line and continuum data. This is effectively required for the study of stellar kinematics, which relies on absorption rather than emission lines.

From the full output cubes we can perform the kinematics analysis in a spatially resolved way to produce 2D kinematics maps for the galaxy's~\vlos~and \vdisp~profiles. The noiseless simulated cubes and the input data are not affected by low SNR issues, allowing us to perform the fit on each individual spaxel. We do however apply a limit of 1 count to the noiseless simulated cubes; spaxels with $<1$ count are not deemed to be realistically detectable. For analysis of the noisy data where the SNR is too low for fitting the spectrum in each spaxel (the typical situation in realistic observations), we apply the Voronoi tessellation method of~\citet{Cappellari2003} to achieve a minimum SNR of 15 in each spatial bin. Even with tessellation the SNR of the true galaxy observed spectral cube is too low for a spatially resolved stellar kinematics analysis; we therefore carry out this portion of the analysis on a simulated high-SNR cube where the flux from the galaxy was increased by a factor of 100.

Where realistic noise estimates are available (i.e. the noisy simulated data), we provide the uncertainties on the fitted parameters from pPXF as 1-$\sigma$ errors. In the following section we will present the results of these analysis steps for the various data products.

\section{Results}\label{sec:results}

\subsection{Input cube analysis}

A major advantage of producing the simulation input data from cosmological simulations is that the input physics and the spectral properties of the galaxy's stellar population are fully known. To study how well we can expect HARMONI to recover the galaxy's stellar kinematics, we first perform the data analysis on the 10 mas-sampled input spectral cube, and on the integrated spectrum derived from it.

\subsubsection{Half-light radius $R_e$}

A key measurement for a distant galaxy is the determination of its half-light radius, and in the following sections we will determine how well this quantity can be measured from IFS observations with HARMONI. To set a reference, the measurement was performed on the input data. From the input spectral cube, the H-band luminosity was determined for each spaxel, thus creating an integrated ``true'' H-band image of the galaxy. Assuming a spherical shape, the radius containing half the galaxy's light was measured to be 0.3 kpc, or 40 mas.

\subsubsection{Integrated spectrum}

We produced an integrated spectrum of the galaxy alongside the spatially resolved spectral cube following the method described in Section~\ref{sec:data_analysis}. We extract the spectrum over the central 0.4\arcsec of the field.

As no uncertainty is available for the template spectra used to create the input cubes, we set the noise to a constant very low value, and quote no uncertainty on the fit parameters. The fit produces an integrated~\vlos~of -48 and~\vdisp~of 180~\kms. These values are uniformly weighted averages over the central 0.4\arcsec~($\sim$3.3 kpc). The negative line-of-sight velocity reflects the global motion of the galaxy, consistent with the simulation output of the \nutfb~simulation.

The results from the integrated spectral fits are summarised in Table~\ref{tab:integrated_summary}.

\begin{figure*}
	\includegraphics[width=\textwidth]{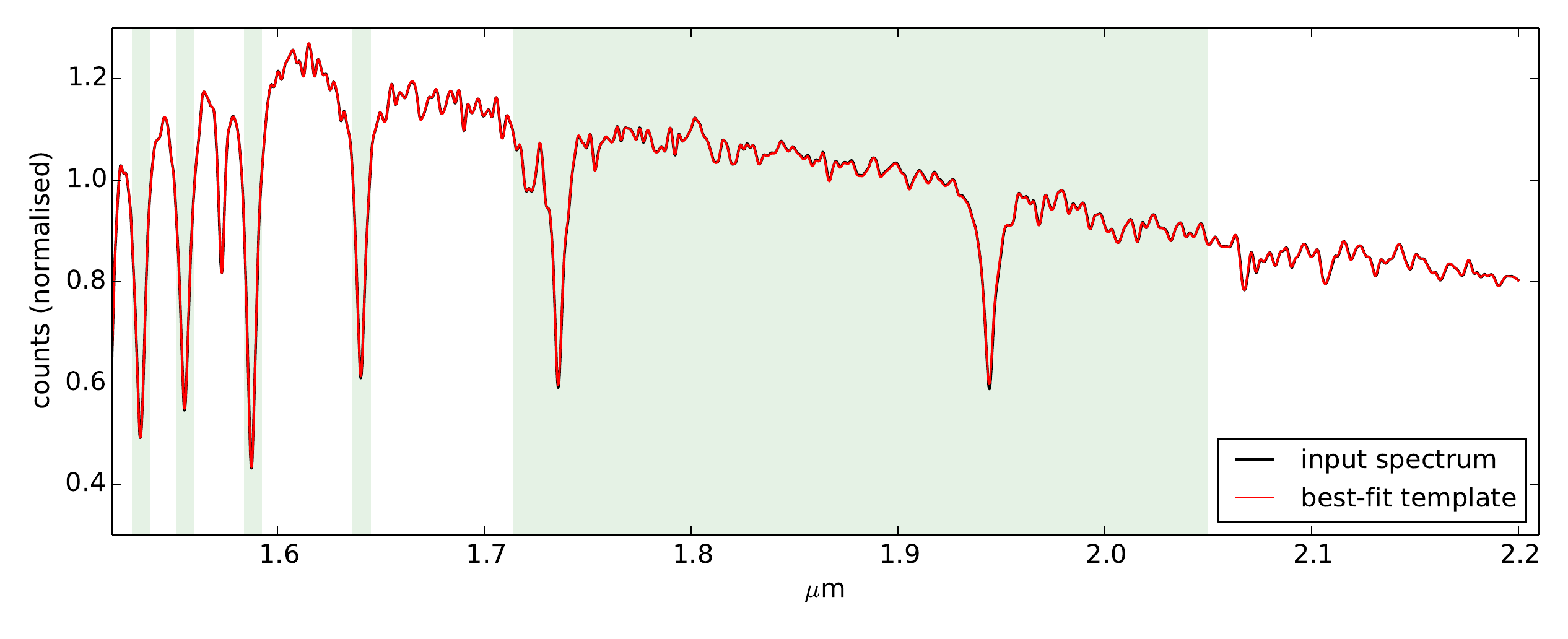}
	\caption{The integrated spectrum calculated from the central 0.4\arcsec~of the input cube (black) together with the best-matching spectral template from the pPXF fitting routine (red). The shaded regions were excluded from the fit: the H absorption lines that may be dominated by emission in a young star-forming system, and the region at $\lambda = 1.72-2.05$~\micron, for which we did not produce an input spectral cube. The blue markers indicate the residuals of the fit, which are in this case very small. The best-fit values for \vlos~and \vdisp~found for this spectrum were 54~\kms and 162~\kms (see Table~\ref{tab:integrated_summary}).}\label{fig:input_int_ppxf}
\end{figure*}

\subsubsection{Kinematics}

The input cube provides the best ``true'' measure of the galaxy's kinematics as traced by its stellar spectra. Various methods for producing v$_{\rm LOS}$ and $\sigma_{*}$ baseline maps from the input cubes are possible. To provide the best fit with what will be measured in our mock observations, we performed the full spectral fitting analysis using pPXF on a spaxel-to-spaxel basis in the 10 mas-sampled input cube. As the star particles' spectra were produced using the same model spectra as used by the fitting routine, each spaxel should in principle be perfectly fit by a linear combination of templates.

\begin{figure*}
	\centering
	\includegraphics[width=13cm]{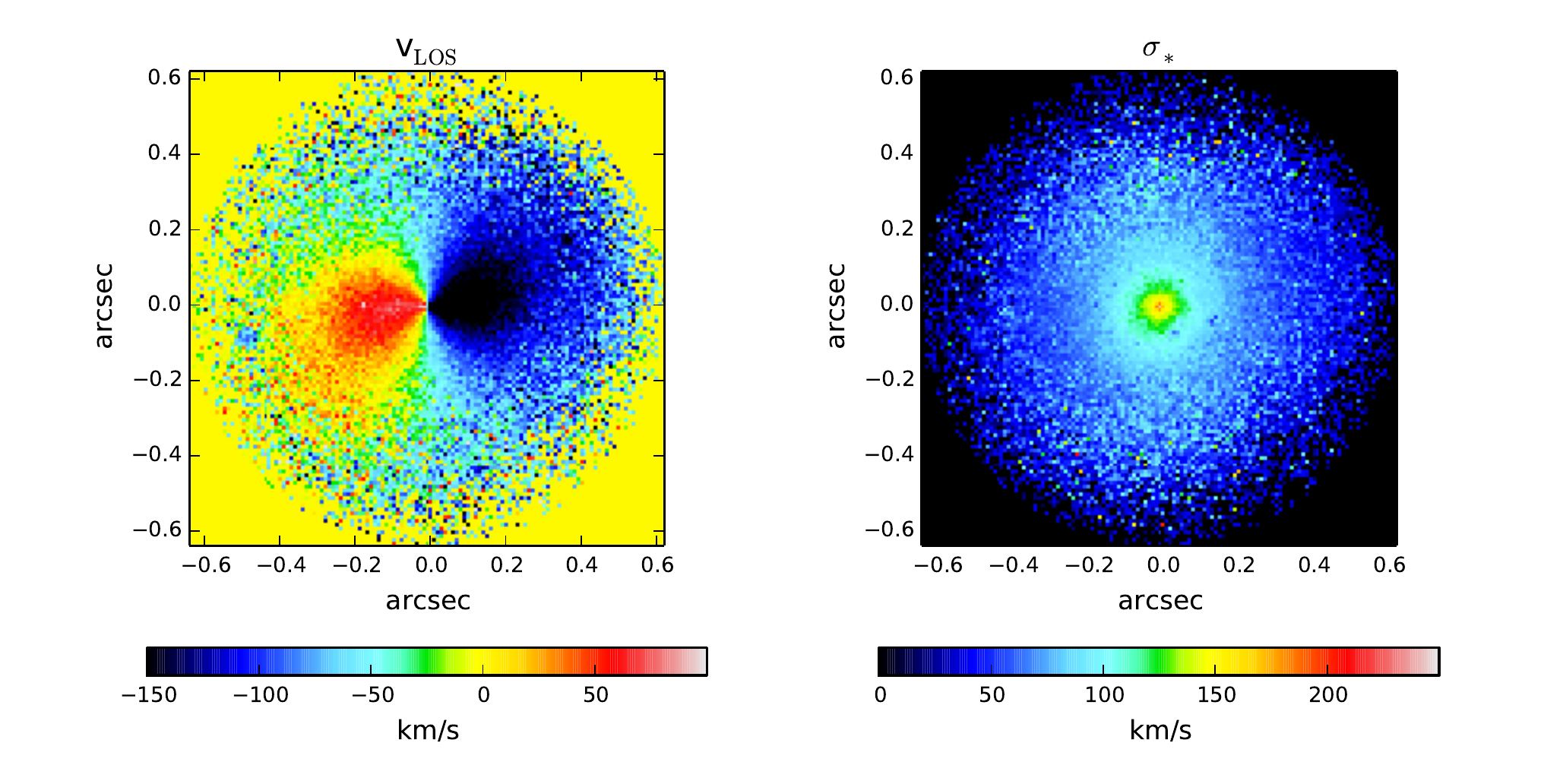}
	\caption{Kinematic maps of the input spectral cube for the HSIM simulations, fit using pPXF in every 10-mas spaxel. (L) Line of sight velocity (v$_{\rm LOS}$); and (R) Velocity dispersion ($\sigma_{*}$). These quantities represent luminosity-weighted averages over the particles in 10 $\times$ 10 mas (80 $\times$ 80 pc) All values shown in~\kms. Typical uncertainties on the fit parameters are $<$0.5~\kms.}\label{fig:input_ppxf}
\end{figure*}

Spaxels with a maximum luminosity below 10$^{-20}$ erg s$^{-1}$ cm$^{-2}$ \AA$^{-1}$ arcsec$^{-2}$, 6 orders of magnitude below the cube's peak values, are considered to produce no detectable light and skipped in the fit. No noise or uncertainty estimate is available for the model spectra. As for the integrated spectrum, we assume a negligibly low noise value to allow the fit to converge; the uncertainty on the fitted kinematics values are $<$ 0.5~\kms, as above. The resulting v$_{\rm LOS}$ and $\sigma_{*}$ maps are shown in Figure~\ref{fig:input_ppxf}. The kinematics maps show a clear signature of ordered disc rotation with a centrally peaked velocity dispersion profile.

\subsection{Noiseless simulated observations}\label{sec:noiseless_results}

\begin{figure*}
 	\centering
 	\includegraphics[width=13cm]{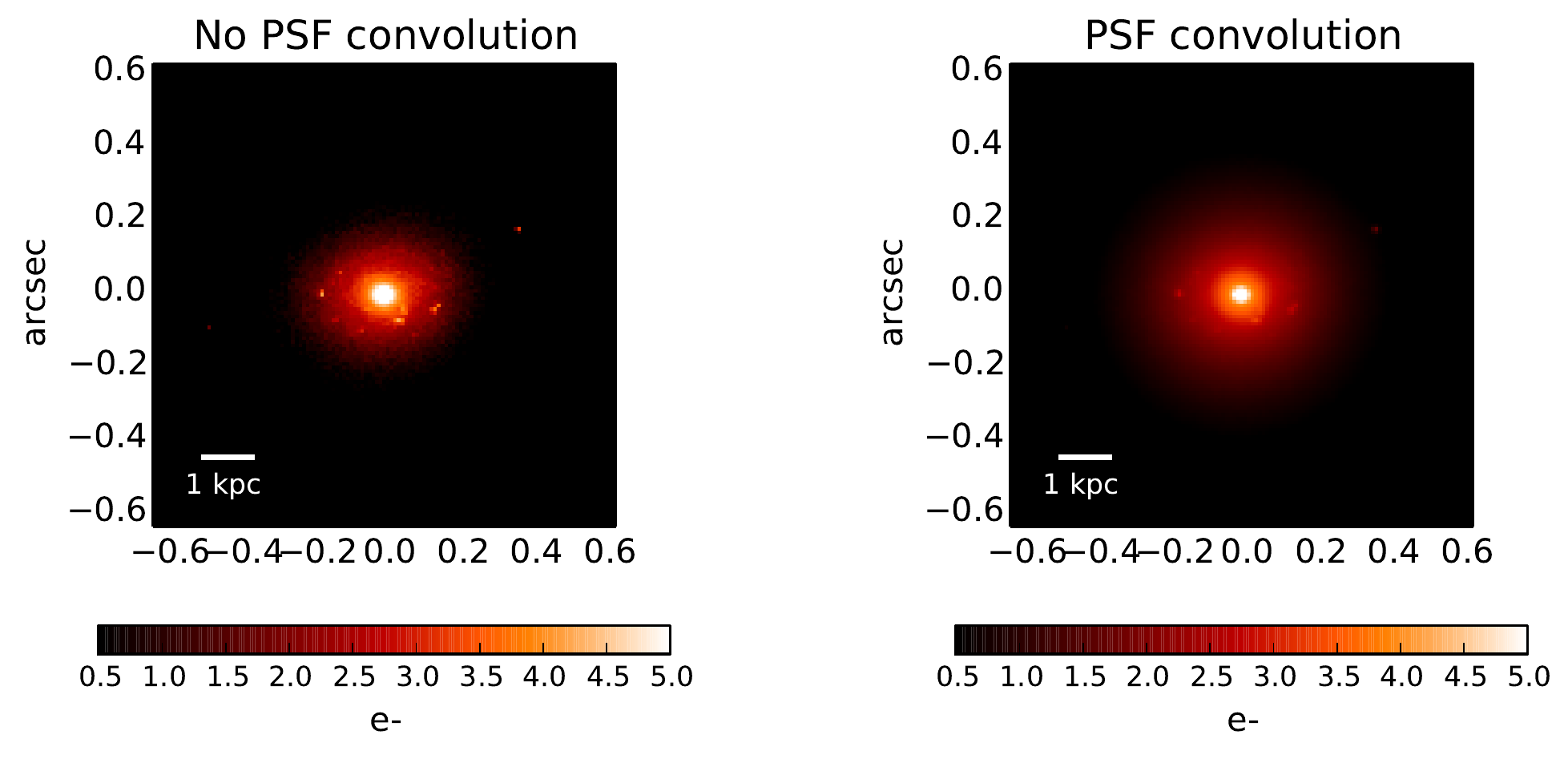}
 	\caption{``Noiseless'' simulated observations of the galaxy, integrated over $\sim$1.5-1.7~\micron~(the Ca H+K region). (L) Without PSF convolution, showing substructure details and faint satellites; and (R) With PSF convolution, showing a much smoother image with less detail.}\label{fig:noiseless_psf}
 	\end{figure*}

As well as the noisy simulated observations, the HSIM pipeline produces noiseless data for both object and background. For the object, the noiseless data represent the detector counts from the source only, transmitted through the atmosphere and telescope and convolved with the AO-corrected PSF. We use this output data here for two reasons: first, to verify the simulation and analysis methods by cross-checking against the input cube; and second, to examine in particular the effect of PSF convolution on the recovered kinematics in the presence of AO correction. Integrated noiseless images with and without PSF convolution are shown in Figure~\ref{fig:noiseless_psf} for the Ca H+K (H-band) spectral region.

\subsubsection{Half-light radius $R_e$}

The noiseless output cube produced without PSF convolution will show the same size and scale as the input cube, but the cubes where PSF convolution was applied offer the opportunity to study the effect of the PSF on the observed galaxy size. A first simple measurement of the observed half-light radius from an integrated image of the noiseless cube yields a value of 0.14\arcsec, corresponding to 1.1 kpc. This is a factor 3-4 higher than the known $R_e$ from the input cube. 

To test whether this degradation can be attributed to the effect of PSF convolution, given our detailed knowledge of the PSF used by HSIM, the galaxy shape of the PSF-convolved image was analysed using GALFIT~\citep{Peng2002,Peng2010}. For the noiseless output cube in the Ca H+K region (1.5-1.7~\micron~, H-band), yielded an $R_e$ value of 4.7 $\pm$ 0.1 spaxels, corresponding to 0.37 $\pm$ 0.01 kpc. This is within 10\% of the value obtained for the input cube, suggesting that PSF convolution is primarily responsible for the broadening of the image in the observation.

\subsubsection{Integrated spectrum}

An integrated spectrum was produced from the noiseless output data following Section~\ref{sec:data_analysis} for the cases with and without PSF convolution; as the field of view is much larger than the physical extent of the galaxy, the integrated spectrum is not expected to be different for both cases. 

Both versions of the noiseless spectra were analysed using pPXF to extract the galaxy's integrated \vlos~and \vdisp. As with the spectrum derived from the input cube, we assume a constant low noise level for error analysis purposes; as this noise is not physically meaningful, we do not quote uncertainties on the fit results. The best-fit~\vlos~and~\vdisp~for the spectrum from the PSF-convolved simulation were found to be -52 and 175~\kms, respectively. These values are consistent with those derived from the input cube. Repeating the analysis on the data without PSF convolution were consistent with the above values, -51 and 175~\kms~for~\vlos~and~\vdisp, respectively. As expected, PSF convolution does not affect the derived integrated kinematics if a large enough aperture can be extracted around the galaxy. 

Results from the fits to integrated spectra are summarised in Table~\ref{tab:integrated_summary}.

\subsubsection{Kinematics}

The spatially resolved kinematics analysis was performed on the noiseless output cubes for both the PSF-convolved and unconvolved cases. The availability of the simulated full spectral cube without PSF convolution allows us to (i) make the ``cleanest'' possible comparison with the velocity maps of the input cube and thus validate the simulation and analysis methods, and (ii) study the effect of the PSF convolution on the retrieval of kinematic properties, in the absence of noise (i.e. for very high S/N). 

The data processing consists of applying the transmission correction to the noiseless output cubes, and stitching the cubes from the Ca H+K and Mg I b spectral regions together. As the noiseless data represent the case of infinitely high S/N, no spatial binning is required to achieve a minimal S/N. As these spectra are given in units of detector counts, i.e. no flux calibration has been applied, we allow pPXF to apply a multiplicative linear polynomial to the spectrum to account for the change in spectral slope as a function of wavelength. The choice of degree for this polynomial was not found to affect the fit results significantly; we note an in-depth investigation of the choice of this parameter in~\citet{vdSande2013}. Spaxels in which the maximum count level over the wavelength range is $<$1 e$^-$ after a 15-hour observation are set to 0 in both \vlos~and~\vdisp~as these are considered un-observable. As for the integrated spectrum and the input cube, noise was set at a constant low value.

\begin{figure}
    \includegraphics[width=9cm]{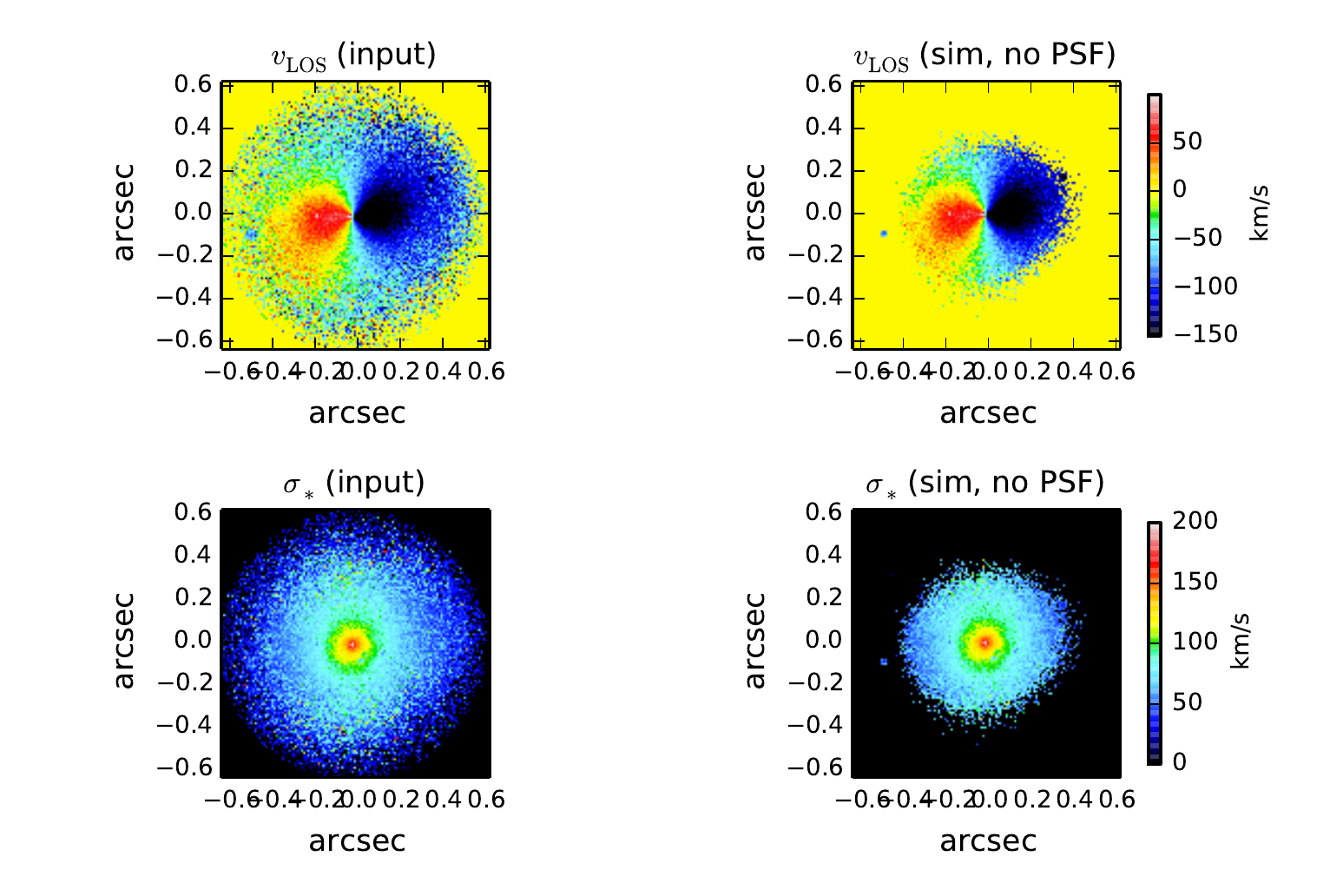}
    \caption{Maps of v$_{\rm LOS}$ (top row) and $\sigma_{*}$ (bottom row) of the input cube (L) and the noiseless simulated spectral cubes without PSF convolution (R) . The simulated maps contain fewer non-zero pixels, as spaxels with $<$ 1 count over the 15-hour (on-source) simulated observation were deemed un-observable and set to zero. The velocity and dispersion maps are plotted on the same scale for both the input and simulated data; the central colour bar applies to both maps in the row. The differencing between input and simulated data excludes the zero-value pixels. Mean values of the difference and standard deviation are labelled in the plots. The figure demonstrates that the kinematics derived from the input and noiseless output cube without PSF convolution are an extremely good match.}\label{fig:noiseless_input_ppxf}
\end{figure}

In Figure~\ref{fig:noiseless_input_ppxf} we show the resulting kinematics maps for the simulations without PSF convolution, which were compared with the corresponding maps from the input cubes (See Figure~\ref{fig:input_ppxf}). The difference between the fitted maps is essentially zero for both quantities. The maximum velocity dispersion at the centre of the galaxy is found to be 253~\kms~for the input cube, and 252~\kms~for the simulated data. From this we conclude that fit results from the noiseless simulated data in the absence of PSF convolution is closely matched to those obtained for the input cube, thus validating our simulation, analysis and fitting processes.

\begin{figure*}
    \includegraphics[width=\textwidth]{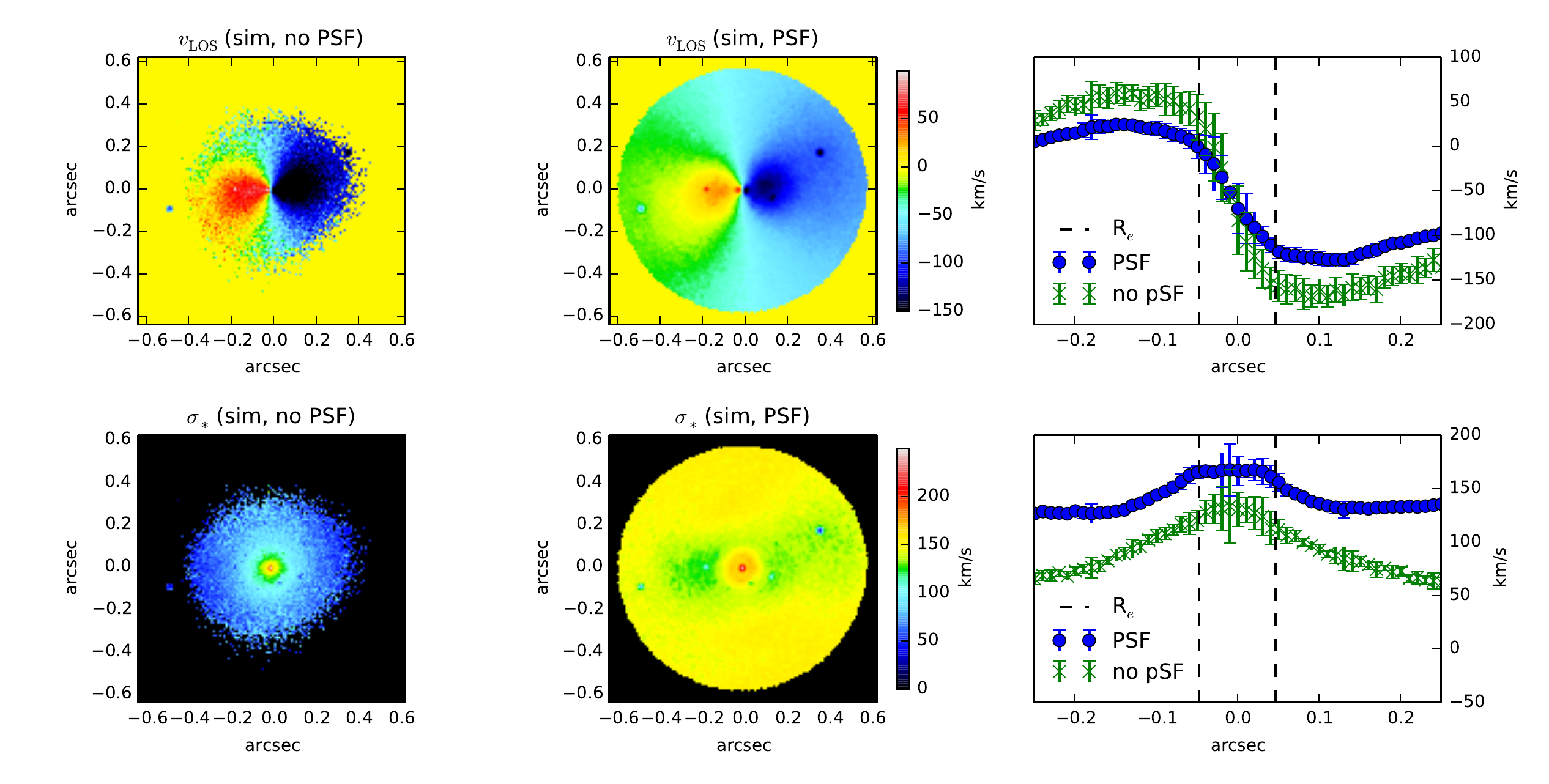}
    \caption{Maps of v$_{\rm LOS}$ (top row) and the uniformly-weighted stellar velocity dispersion $\sigma_{*}$ in each 10 mas spaxel (bottom row) for the noiseless simulated spectral cubes of the $z=3$ galaxy, comparing the simulation results with (middle) and without (left) PSF convolution. During the fit, spaxels where the maximum counts do not reach 1 over the observation are excluded from the fit, as before; for the PSF-convolved data this cut-off is much larger and more clearly defined due to the geometric shape of the PSF used in the simulation. The right panels show 2-dimensional slices of~\vlos~vs. radius, showing the effect of beam smearing on the recovered rotation curves for the idealised noiseless data cubes. The data points represent the median~\vlos~and~\vdisp~values in a 1 kpc-wide slit (13 pixels) along the major axis of the galaxy as determined by GALFIT. The errorbars show the standard deviation of the values across the 1 kpc aperture at each pixel along the axis; as the data shown are noiseless we assume all other sources of uncertainty to be negligible. The galaxy's half-light radius R$_e$ is indicated with the dashed line. Note that the turnover of the rotation curve in the galaxy's out regions is a result of the unrealistic stellar mass to halo mass ratio of the galaxy.}\label{fig:noiseless_kinematics}
\end{figure*}

To study the effect of PSF convolution, the PSF-convolved spectral cube was then fitted in the same way, using pPXF on a spaxel-to-spaxel basis. Again velocity and dispersion for the spaxels where the maximum count level is $<$ 1 for the 15-hour on-source exposure are set to zero, and the same input parameters are used for the fitting procedure. The resulting maps are shown in Figure~\ref{fig:noiseless_kinematics} alongside those derived from the no-PSF simulation. In the Figure we also plot~\vlos~as a function of radius in the galaxy for the inner 0.5\arcsec~(4 kpc; approx. 10 R$_{e}$). These values were measured along a 1 kpc-wide (13 pixels) rectangular aperture along the galaxy's major axis, the position angle of which was taken from the GALFIT results. The angle is very small at just 2\degree. The data points show the median of the 13 velocity or velocity dispersion measurements at each radial position; the error bars represent the standard deviation. As the data are noiseless, we assume all other sources of uncertainty are negligible. The size of the half-light radius R$_e$ is indicated with the dashed line.

The fitted region is much larger in the convolved data, carrying the signature of the PSF, which spreads the galaxy's starlight over a larger area. As expected the slope of the velocity curve is flatter due to the smearing effects of the PSF; the dispersion is higher but does not fall off as steeply as in the no-PSF case. As reported in~\citet{Davies2011} the very central velocity dispersion is in fact higher in the no-PSF case, however as the rotation curve falls off more steeply and we median over a relatively wider aperture, the PSF-convoled case appears to give higher values along the entire axis. Our results are therefore consistent with other studies for seeing-limited observations, and indicate that even in the presence of advanced AO facilities in the NIR, PSF convolution can considerably complicate the interpretation galaxy dynamics from IFU observations. This will be discussed in more detail in Section~\ref{sec:discussion}.

\subsection{Low S/N simulated observations}

In this section we extend the analyses shown above to the ``true'' simulated data, which include the effects of source, background and detector noise, over the 15-hour observation. For the initial data processing, sky subtraction was performed, as well as the correction for atmospheric transmission and telescope and instrument throughput, as would be done using a standard star observation in a real observation. The data products required for these processing steps are returned by the HSIM pipeline. A noise cube was created from the difference of the noisy and noiseless output data; we carry this through the analysis procedures described below to obtain realistic uncertainty estimates on the fits.

\subsubsection{Half-light radius $R_e$}

As for the noiseless data, we analyse the noisy data using the GALFIT code to derive the galaxy's half-light radius $R_e$. Using the Ca H+K output datacube and an H-band LTAO-corrected PSF, we find an $R_e$ of 4.3 $\pm$ 0.4 spaxels, corresponding to 0.34 $\pm$ 0.03 kpc. This is in agreement with the value retrieved from the noiseless data to within 2-$\sigma$. 

\subsubsection{Integrated spectrum}

\begin{figure}
    \includegraphics[width=8cm]{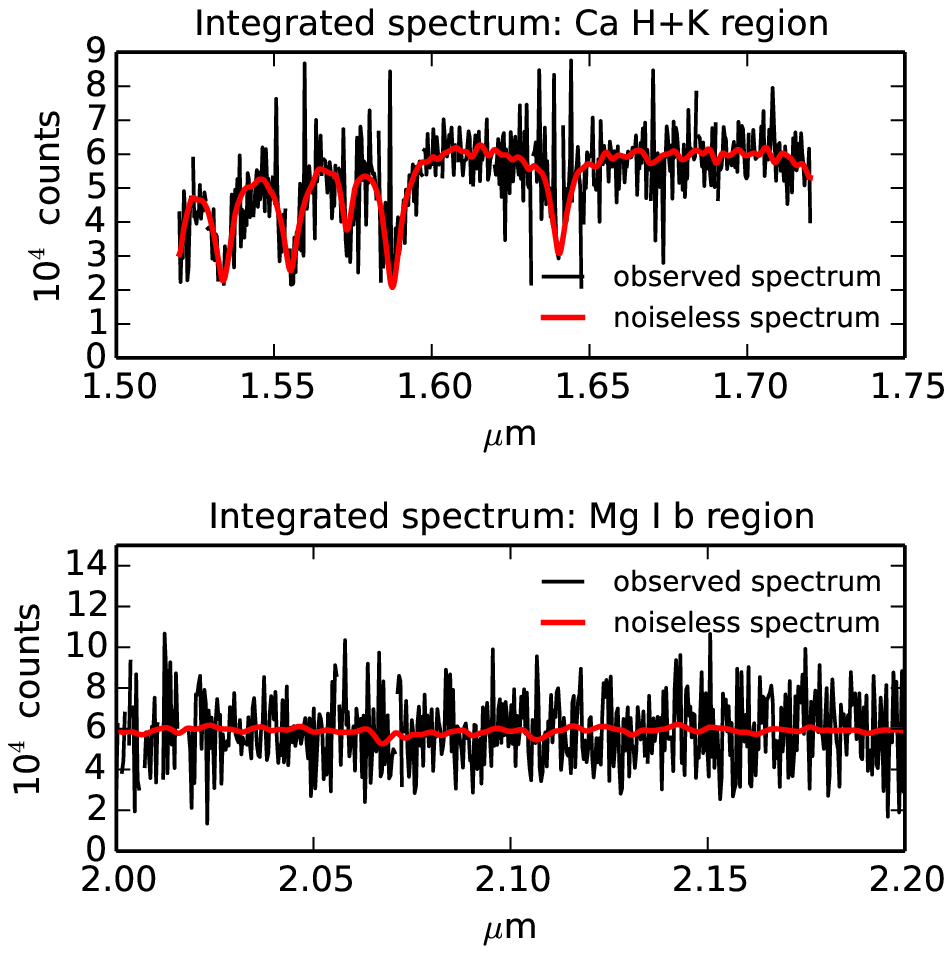}
    \caption{Integrated spectrum extracted from the central 0.43\arcsec~of the field of view, centred on the $z=3$ galaxy, for the two spectral regions considered in this study. The spectra were clipped at 3-$\sigma$. The corresponding noiseless spectrum is overplotted in red in each panel.}\label{fig:noisy_spectrum}
\end{figure}

As discussed above, producing an integrated spectrum of the galaxy from the full field of the noisy output cube will add too much noise to the spectrum when using a simple co-adding procedure. For this study, we limit the noise by focusing on the central 0.43\arcsec~of the field, and co-add the spaxel spectra to produce the integrated spectrum. Extracted over this aperture, the average SNR over the full wavelength range is 7. The (PSF-convolved) spectrum extracted from this central region is plotted in Figure~\ref{fig:noisy_spectrum} together with the corresponding noiseless spectrum showing the ``clean'' object signal.

\begin{figure*}
    \includegraphics[width=\textwidth]{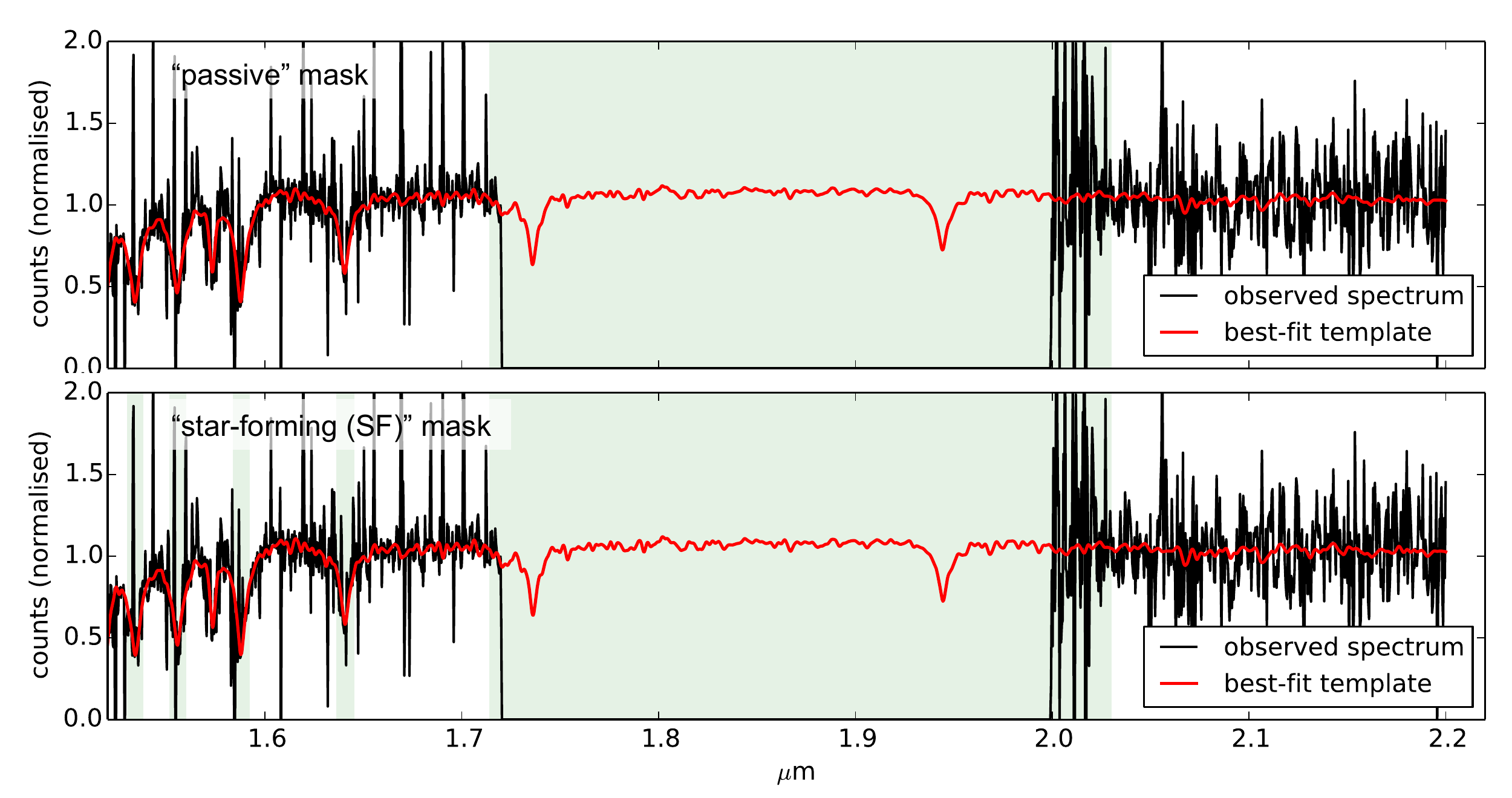}
    \caption{Integrated noisy spectrum extracted from the central 0.4\arcsec~of the field of view of the simulation with PSF convolution (black), overlaid with the best-fit template (red). The shaded regions indicate those excluded from the fit. We show both the ``passive'' (top) and ``active'' (i.e. star-forming; bottom) cases. In the latter, H absorption lines may be filled in with emission and thus unusable for absorption line fitting. The availability of multiple strong absorption lines improved the recovery of the galaxy's input kinematics markedly for these low S/N spectra. The fit results are listed in Table~\ref{tab:integrated_summary}.}\label{fig:noisy_spectrum_ppxf}
\end{figure*}

The spectral fitting procedure using pPXF was applied to the noisy spectra extracted from the central 0.43\arcsec~region. As realistic noise values are available for this spectrum, we report the uncertainties on the fit parameters as returned by pPXF. We find best-fit values for~\vlos~and~\vdisp~of 4 $\pm$ 26~\kms~and 126 $\pm$ 33~\kms, respectively. Despite a good-quality fit (see Figure~\ref{fig:noisy_spectrum_ppxf}), these values are discrepant from the input values. We note however that with the H lines masked out from the fit to mimic a star-forming (``SF'') galaxy, just one strong line (Ca K) is available for the kinematic fit. To investigate the effect of this on the outcome, we repeat the fit allowing pPXF to use the H absorption lines, as would be possible for a passive system. In this case we find \vlos~and \vdisp~of -35 $\pm$ 20~\kms~and 140 $\pm$ 32~\kms; within the error bars these values are consistent with the input. This demonstrates the importance of having good spectral indicators available for fitting when the S/N is low. For star-forming galaxies and AGN in particular this could place constraints on the redshift ranges most suitable for stellar kinematics and stellar population studies.

For the case where the H absorption lines are masked the difference in velocity dispersion is approximately 40\%. Given that a galaxy's dynamical mass $M_{\rm dyn} \propto \sigma_{*}^2$, this measurement would lead to a factor 2 difference in the calculated dynamical mass.

The spectrum with its best-fit template is shown in Figure~\ref{fig:noisy_spectrum_ppxf}, for both line masking cases. The spectra extracted from runs without PSF convolution were consistent with the PSF-convolved observations, we do not show or list these results separately. 

\subsection{High S/N simulated observations}

The~\nutfb~galaxy's integrated spectrum after a 15-hour (on-source time) observation with HARMONI on the E-ELT has a SNR of $<$10, with only the  brightest central spaxels showing a SNR of $>$ 3. We showed that this is sufficient for reliably deriving kinematics parameters, however, it is too low for producing a spatially resolved kinematics map, even when spatially binning with the Voronoi tessellation to obtain higher S/N. We comment further on this issue in the context of high-z galaxy studies with HARMONI in Section~\ref{sec:discussion}.

To perform a quasi-realistic kinematics analysis of the galaxy in the presence of noise, we performed an additional set of simulations with the galaxy's flux brightened by a factor 100; these results are shown in the following section. The data used for this part of the analysis are identical to the original input cube of the~\nutfb~galaxy, the only difference is the 100 times higher flux. The HSIM parameters were the same as in the low-SNR simulations. 

We first perform the kinematics fit on the integrated spectrum, which was produced as outlined in Section~\ref{sec:data_analysis}, paying particular attention to the effect of different masking schemes to mimick passive or star-forming systems. We find that for the high S/N case, the availability of spectra features has less effect on the outcome of the fit, with both fits giving results in agreement with each other and with the input values (see Table~\ref{tab:integrated_summary}).

We then use the high-SNR data for the spatially resolved kinematics analysis. As above, we focus our analysis on the central 0.4\arcsec~of the full field, corresponding approximately to a 5 R$_e$ radius.

\begin{figure*}
    \includegraphics[width=\textwidth]{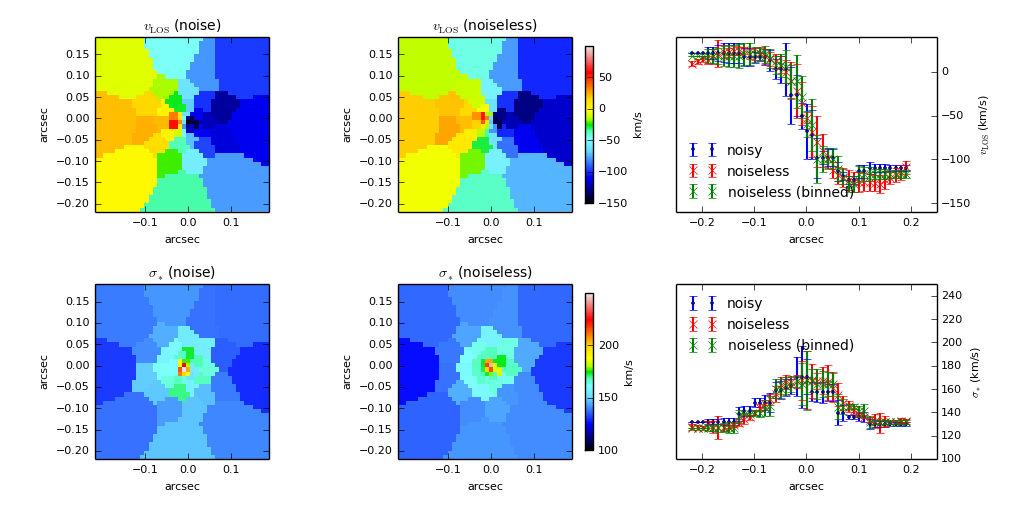}
    \caption{Velocity and stellar velocity dispersion maps of the~\nutfb~galaxy including realistic noise, based on the 15-hr observation of the galaxy brightened by a factor 100 to improve the SNR. The data were spatially binned to achieve a SNR of $>$15 in all bins. The top row panels show the~\vlos~values for the noisy data (left), and for the corresponding noiseless observations, binned using the same Voronoi scheme (middle). The colour bar applies to both panels for ease of comparison. The right-hand panel shows the rotation curve derived in a 1-kpc wide rectangular aperture along the galaxy's major axis, comparing the noisy measurements with the spatially binned noiseless data, and the original (unbinned) noiseless data. Bottom row: As the top row, for the stellar velocity dispersion measurements.}\label{fig:noisy_kinematics}
\end{figure*}

The~\nutfb~galaxy with higher flux level produces very high SNR in the galaxy centre when observed with HARMONI on-source for 15 hours; however the drop-off in signal outside the central kpc region remains steep given the galaxy's compact morphology. Spatial binning is required to achieve sufficient SNR in the outer regions for a kinematics analysis. We perform a Voronoi binning of the field to achieve a minimum SNR of 15; note that the high SNR at the galaxy centre provides a number of single-spaxel bins, giving a very high effective spatial resolution in the inner regions. For each bin, a spectrum was produced by unweighted addition of the individual spaxel spectra, and a corresponding noise spectrum was produced by subtracting the corresponding noiseless signal. Full spectral fitting was performed with pPXF to obtain the velocity and velocity dispersion in each bin. Results are shown in Figure~\ref{fig:noisy_kinematics} and compared with the kinematics maps produced from noiseless data, binned spatially following the same Voronoi scheme as used for the noisy data, to ensure an accurate comparison.

The figure shows a good agreement between the noisy and noiseless simulations, indicating that the pPXF fits to spectra with a S/N of 15 provide excellent agreement with the input datacube. Similarly, comparison of binned versus unbinned rotation curves shows that we can recover rotation curves and kinematic maps with a high degree of accuracy. We note that our artificially brightened galaxy represents a very high S/N case that is not fully representative of a typical $z=3$ galaxy; we discuss this limitation in Section~\ref{sec:obscontext}.

\begin{table*}
    \centering
    \begin{tabular}{|l|c|c|c|c|}
        \hline
        Spectrum & PSF convolution & Masking & \vlos (\kms) & \vdisp (\kms)\\
        \hline
        Input & No & SF & -53.7 & 161.8 \\
        Noiseless output & No & SF & -51.7 & 175.0 \\
        Noiseless output & Yes & SF & -49.7 & 174.1\\
        Noisy output & Yes & SF & 4.4 $\pm$ 25.7 & 125.5 $\pm$ 32.8\\
        Noisy output & Yes & Passive & -35.2 $\pm$ 19.7 & 139.8 $\pm$ 31.8\\
        Noisy output (bright) & Yes & SF & -50.1 $\pm$ 1.0 & 174.6 $\pm$ 1.0\\  
        \hline
    \end{tabular}
    \caption{Summary of integrated kinematic values derived from the input and output integrated spectra. All values quoted are in \kms, and obtained from the spectra extracted in the central 0.4\arcsec~of the field. For the input and noiseless data, we do not quote uncertainties as no noise is available for these spectra, and the fit uncertainties are therefore unphysical. The masking column indicates the spectral lines that were available for the fit procedure: ``SF'' (star-forming) is the pessimistic case, where the stellar H absorption lines are entirely filled in with emission, for the ``passive'' case all H lines are available for the fit. The results show that the ``passive'' fits reproduce the input values better in the low S/N case.}\label{tab:integrated_summary}
\end{table*}

\section{Discussion}\label{sec:discussion}

\subsection{The~\nutfb~galaxy in observational context}\label{sec:obscontext}

The~\nutfb~galaxy provides a convenient target for our simulations, given the high spatial resolution of the simulation and its redshift, which lies within our region of interest for this study. Furthermore, the input physics and assumptions of the simulation, and the physical properties of the star particles, are perfectly known, allowing us to understand how well we can recover morphological and kinematic properties from our simulated observations. In this section, we examine how representative the galaxy is of the observed population at high redshift, and the implications for future observations with HARMONI.

Whilst spatially resolved information of galaxies at $z=3$ is limited with present-day instrumentation, some physical properties of the population at these redshifts are reasonably well constrained from imaging and integrated spectroscopic observations. Our galaxy deviates from these in a number of ways. The star formation and feedback prescriptions employed in the~\nutfb~simulation have limitations that translate into discrepancies between its total stellar mass and the stars' spatial distribution from observed systems. 

Formation of a star particle is triggered when a gas particle reaches a critical density $n_{th}$ (see Table~\ref{tab:input}), with a fixed star formation efficiency per free-fall time and strictly following a Schmidt law. Whilst the Schmidt law accurately represents the relation between star formation rate and gas density on a global scale, observations show the relation to be dependent on local environment, and star formation efficiencies are seen to vary strongly within galaxies~\citep{Bigiel2008}.

Second, the role of feedback is critical in the regulation of star formation over cosmic time~\citep{Katz1996}. The \nutfb~simulation includes the effect of supernova feedback; however as in many simulations the feedback is relatively inefficient~\citep{Dubois2008, Kimm2011} and fails to slow the star formation rate at early cosmic times. As a result, the~\nutfb~galaxy displays the typical properties associated with overcooling: a very compact morphology with little sign of an extended stellar disk, and stellar mass fraction inconsistent with observations. The turnover of the stellar rotation curve measured from the kinematic maps shown in Figure~\ref{fig:noiseless_kinematics} is also a result of this inconsistency of the stellar mass in relation to the halo mass.

In their derivation of the stellar mass halo mass (SMHM) relation,~\citet{Behroozi2013} suggest a stellar mass fraction of $<$ 1\% at $z=3$, including contributions from satellite galaxies within the halo, with the caveat that the relation is poorly constrained by observations in this mass-redshift regime. With a central galaxy stellar mass of $\sim$10$^{10}$\msun, not including the masses of the 2 low-mass satellites, the~\nutfb~galaxy's stellar mass lies approximately an order of magnitude above the SMHM relation; or conversely, a 10$^{10}$~\msun~stellar mass galaxy is likely to reside in a 10$^{12}$~\msun~halo at this redshift. Detailed studies into implementation schemes for supernova and/or massive stellar feedback have been able to produce galaxies with more realistic star formation histories, e.g.~\citet{Kimm2015, Agertz2015, Hopkins2011}.

Our S/N calculations from both Section~\ref{sec:ryan} and the simulation results shown in Section~\ref{sec:results} are consistent in predicting that a 10$^{10}$~\msun~galaxy at $z=3$ is likely to be too faint for spatially resolved absorption line integral field spectroscopy with HARMONI. In Section~\ref{sec:ryan} we estimate a limit of $\sim 10^{10.7}$~\msun~for \emph{passive} galaxies.

In Figure~\ref{fig:noisy_kinematics} we show the accurate recovery of the galaxy's intrinsic stellar kinematics when its luminosity was increased by a factor of 10$^2$, corresponding to  brightening of 5 magnitudes ($M_V \sim -26$). In the context of the V-band galaxy luminosity function~\citep{Marchesini2012} at $z=3$ this makes the galaxy no longer representative of the known population at that redshift ($\sim 30 L_*$). We note however that the bright end of the luminosity function at high redshift is poorly constrained, in part due to the relatively small areas probed by today's surveys. HARMONI will operate concurrently with the Large Synoptic Survey Telescope~\citep[LSST;][]{Ivezic2008}, which will provide photometry for 10$^{10}$ galaxies. It will be capable of detecting Lyman-break galaxies with $L > L_*$ to $z > 5.5$ and passively evolving galaxies with with $L > L_*$ to $z > 2$ over 20,000 deg$^2$, and go several magnitudes deeper for smaller fields~\citep{LSST_book}.

However our results do confirm that spatially resolved absorption line spectroscopy will be within reach to $z=3$ at the bright end of the luminosity function with realistic exposure times. Such observations will allow us to probe the dynamics and stellar population characteristics (e.g. ages, metallicities) at this critical era in galaxy evolution. Furthermore, with the range of spatial and spectral settings available in HARMONI, observations can be tailored in resolution and sensitivity to  range of galaxy properties; our results show that for the brightest most massive galaxies, the cores of even very compact galaxies can be spatially resolved.

\subsection{Integral field spectroscopic studies with advanced AO systems}

The advent of AO has greatly improved the spatial resolution achievable with IFS on 8-10-m telescopes, allowing us to probe physical scales to $\sim$1 kpc in galaxies at $z >1$, compared with $\sim$5 kpc for seeing limited observations. With the E-ELT and HARMONI, we will be able to probe to around 200-500 pc in this redshift regime - the scale of individual giant HII regions. On the 39-m E-ELT and its ELT-class counterparts, advanced AO systems are required to fully exploit the telescope's aperture, and ensure good sky coverage. However the resulting AO-corrected PSF is more complex than that for a single-conjugate AO system on today's 10-m-class telescopes, or that of a seeing-limited observation, and its impact on IFS observations must be well understood to place the recovered kinematic parameters into physical context. 

\citet{Zieleniewski2013} describe the method employed to produce realistic PSF models of LTAO-assisted E-ELT observations over the full wavelength range covered by HARMONI for inclusion in HSIM. Though the effect of S/N, spatial resolution and beam smearing are relatively well studied, the complexity of the LTAO PSF warrants revisiting this issue for next-generation IFS instruments.

The results we present in Section~\ref{sec:noiseless_results} for the noiseless simulated observations illustrate very clearly the effect of the LTAO PSF convolution. Particularly marked is the difference between observed an intrinsic and observed half-light radius $R_e$, which suggests the ensquared energy radius increases by a factor of 3-4 even for near-diffraction-limited image quality. We do however note that given knowledge of the PSF, the intrinsic galaxy shape and $R_e$ was recovered with commonly used fitting tools such as GALFIT even for low SNR observations.

For the spatially resolved kinematics, beam smearing flattens the slope of the rotation curve and increases the measured velocity dispersion (see Figure~\ref{fig:noiseless_kinematics}), which must be taken into account when extracting derived quantities such as the galaxy's dynamical mass. A number of correction methods have been developed to account for this, e.g. in~\citet{Burkert2015, Green2010, Cresci2009}, and we refer to~\citet{Davies2011} for a comprehensive review and comparative study of the different methods and their biases for a range of different weighting methods at low and high S/N (seeing-limited) observations. There is however no single optimal method to account for the effects of beam smearing and spatial resolution, and further modelling for HARMONI in difference spatial sampling and S/N regimes is warranted.

The data products produced by HSIM make it a suitable tool for studying this effect and test recipes for accurately accounting for the effects of beam smearing when deriving a galaxy's properties from IFS data. This is particularly valuable given that in the E-ELT era, no images will be available for HARMONI's targets with higher spatial resolution, as is currently possible via space-based imaging with the Hubble Space Telescope.

\subsection{Beyond the proof-of-concept study}

The simulations presented here form a proof-of-concept study into the use of cosmologial simulations as input for the HARMONI instrument simulator, HSIM. Present-day cosmological simulations provide hundreds of thousands of galaxies throughout cosmic time. With the agreement between observations and simulations improving greatly in recent years, mock observations of simulated galaxies are a valuable tool for characterising the performance of next-generation instruments in currently inaccessible areas of parameter space - with the added advantage that the input physics and the galaxies' physical properties are a priori perfectly known. Our work illustrates this, using a single galaxy from the high-resolution \nutfb~simulation.

For this initial study we have used the HSIM simulation pipeline described in ~\citet{Zieleniewski2015}, and developed additional tools to convert RAMSES star particle data into full spectral cubes of the galaxy's starlight suitable as HSIM input, as well as post-processing to implement quasi-realistic data analysis procedures. A major future goal of this work is to model galaxies' gas emission to obtain fully realistic spectral cubes for passive as well as star-forming galaxies.

As a whole, this pipeline can be generalised and used for a range of purposes. From an instrumental point of view, producing mock observations of simulated galaxies provides physically realistic data for studying observational biases and data analysis methods, and for exploring the instrument's capabilities for a range of targets. Second, it also allows us to make observational predictions from the wealth of data produced in present-day cosmological simulations, and use this data to design observational surveys with the next generation of observational facilities.

\section{Conclusions}\label{sec:conclusions}

We have presented the simulation tools developed to simulate observations of realistic star-forming and passive galaxies at $2 < z < 4$ with the E-ELT integral field spectrograph HARMONI, using as input first simple analytical light profiles, and more complex models from a high-resolution cosmological simulation. From the results of these studies we summarise our conclusions as follows.

\begin{itemize}
	\item HSIM simulations of model \emph{passive} galaxies indicate that integrated spectroscopy of these galaxies at $2 < z < 4$ will be limited to stellar masses $\gtrsim 10^{10.7}$~\msun~for 10 hours of on-source exposure time. Star-forming galaxies are intrinsically brighter and may be observable to lower stellar masses. We note here the importance of strong absorption lines for fitting of stellar kinematics, of which there may be fewer in galaxies that are actively forming stars. 
	
	\item Cosmological simulations provide useful data for creating model galaxies with more realistic morphologies, dynamical profiles, and star formation histories. For this work, we have developed the tools to convert the output of one such simulation, \nutfb, into a 3-dimensional spectral cube. These software tools will allow us to produce a larger sample of model galaxies, which, in combination with the HSIM pipeline, can be mock-observed with HARMONI.
	
	\item A mock 15-hour (on-source) observation of the main \emph{star-forming} galaxy from the~\nutfb~simulation at $z=3$, with a stellar mass of 10$^{10}$~\msun ($\sim$ 0.3 L$_*$ at $z=3$), indicates that HARMONI can achieve a sufficiently high S/N for integrated spectroscopy to recover the~\vlos~and~\vdisp~using common spectral fitting tools such as pPXF. Our analysis does however show that the availability of strong absorption features is important for the quality of the stellar kinematics fit; this could limit the accuracy of such studies for star-forming or active galaxies at high redshift. The S/N is not sufficient for spatially resolved spectroscopy. This is consistent with the results from our simulated observations of simple galaxy models. We note that given the current limiting redshift for absorption line IFS studies ($z < 0.2$), probing the stellar kinematics and population characteristics with HARMONI even at $z > 1$ will be a major advance.
	
	\item When we artificially brighten the~\nutfb~galaxy by a factor of 100 ($\sim$ 30 L$_*$), the stellar kinematics and the rotation curve are well recovered to beyond 1 $R_e$. Whilst galaxies of this luminosity do not represent the ``typical'' population at this redshift, we note that the large-scale surveys in the E-ELT era, in particular LSST, will cover a vastly larger area of sky than currently available. This is particularly valuable for discovering galaxies at the bright end of the luminosity function. It is therefore likely that HARMONI will be able to produce spatially resolved IFS of a small number of galaxies at $z=3$.
	
	\item In addition to providing performance estimates for HARMONI, our simulations also show how HSIM can be used to study observational biases and how they affect the kinematics recovered from the spectra after data reduction and analysis. The effects of PSF convolution in particular, which are relatively well understood for seeing limited observations, may well require more detailed modelling to investigate the effect of the more complex PSF shape resulting from LTAO correction. Comparison of our simulations with and without PSF convolution indicate that the kinematics are significantly affected by beam smearing due to the broad PSF halo, despite the image quality being nominally diffraction-limited in the NIR. Critically in the E-ELT era no higher-resolution imaging will be available to correct for PSF effects, as is currently possible through space-based imaging; sophisticated modelling tools like HSIM are therefore highly valuable in understanding such effects.
	
	\end{itemize}

This study demonstrates the potential of combining high-resolution cosmological simulations with sophisticated instrument simulators to optimise the scientific return from future observational facilities: on the one hand allowing us to predict the parameter space accessible to the instrument, plan sample selections and observational strategies ; and make meaningful observational predictions from galaxies in a simulated Universe with perfectly known physics on the other. 

\section*{Acknowledgments}

SK, SZ, RH, NT, MT, FC and KOB are partly or wholly supported by the Science and Technology Facilities Council (STFC) grants ST/J002216/1 and ST/M007650/1, part of the UK E-ELT Programme at the University of Oxford.  These authors also acknowledge support from the ESO HARMONI Interim Study 51921/13/54852/HNE. RCWH was supported by STFC under grant numbers ST/H002456/1 and ST/K00106X/1. The \nutfb~simulation used in this paper was run on the DiRAC facility, jointly funded by BIS and STFC. The research of JD is supported by A. Beecroft, the Oxford Martin School
and STFC. We thank the anonymous referee for their constructive comments and suggestions. SK made extensive use of NASA's Astrophysics Data System Bibliographic Services for this work, as well as the Astropy and pynbody Python packages~\citep{astropy, pynbody}.

\bibliographystyle{mnras}
\bibliography{harmoni-ramses}

\begin{thebibliography}{}
\makeatletter
\relax
\def\mn@urlcharsother{\let\do\@makeother \do\$\do\&\do\#\do\^\do\_\do\%\do\~}
\def\mn@doi{\begingroup\mn@urlcharsother \@ifnextchar [ {\mn@doi@}
  {\mn@doi@[]}}
\def\mn@doi@[#1]#2{\def\@tempa{#1}\ifx\@tempa\@empty \href
  {http://dx.doi.org/#2} {doi:#2}\else \href {http://dx.doi.org/#2} {#1}\fi
  \endgroup}
\def\mn@eprint#1#2{\mn@eprint@#1:#2::\@nil}
\def\mn@eprint@arXiv#1{\href {http://arxiv.org/abs/#1} {{\tt arXiv:#1}}}
\def\mn@eprint@dblp#1{\href {http://dblp.uni-trier.de/rec/bibtex/#1.xml}
  {dblp:#1}}
\def\mn@eprint@#1:#2:#3:#4\@nil{\def\@tempa {#1}\def\@tempb {#2}\def\@tempc
  {#3}\ifx \@tempc \@empty \let \@tempc \@tempb \let \@tempb \@tempa \fi \ifx
  \@tempb \@empty \def\@tempb {arXiv}\fi \@ifundefined
  {mn@eprint@\@tempb}{\@tempb:\@tempc}{\expandafter \expandafter \csname
  mn@eprint@\@tempb\endcsname \expandafter{\@tempc}}}

\bibitem[\protect\citeauthoryear{{Agertz} \& {Kravtsov}}{{Agertz} \&
  {Kravtsov}}{2015}]{Agertz2015}
{Agertz} O.,  {Kravtsov} A.~V.,  2015, preprint, \href
  {http://adsabs.harvard.edu/abs/2015arXiv150900853A} {} (\mn@eprint {arXiv}
  {1509.00853})

\bibitem[\protect\citeauthoryear{{Astropy Collaboration} et~al.,}{{Astropy
  Collaboration} et~al.}{2013}]{astropy}
{Astropy Collaboration} et~al., 2013, \mn@doi [\aap]
  {10.1051/0004-6361/201322068}, \href
  {http://adsabs.harvard.edu/abs/2013A%26A...558A..33A} {558, A33}

\bibitem[\protect\citeauthoryear{{Bacon} et~al.,}{{Bacon}
  et~al.}{2001}]{Bacon2001}
{Bacon} R.,  et~al., 2001, \mn@doi [\mnras] {10.1046/j.1365-8711.2001.04612.x},
  \href {http://adsabs.harvard.edu/abs/2001MNRAS.326...23B} {326, 23}

\bibitem[\protect\citeauthoryear{{Behroozi}, {Conroy}  \&
  {Wechsler}}{{Behroozi} et~al.}{2010}]{Behroozi2010}
{Behroozi} P.~S.,  {Conroy} C.,   {Wechsler} R.~H.,  2010, \mn@doi [\apj]
  {10.1088/0004-637X/717/1/379}, \href
  {http://adsabs.harvard.edu/abs/2010ApJ...717..379B} {717, 379}

\bibitem[\protect\citeauthoryear{{Behroozi}, {Wechsler}  \&
  {Conroy}}{{Behroozi} et~al.}{2013}]{Behroozi2013}
{Behroozi} P.~S.,  {Wechsler} R.~H.,   {Conroy} C.,  2013, \mn@doi [\apj]
  {10.1088/0004-637X/770/1/57}, \href
  {http://adsabs.harvard.edu/abs/2013ApJ...770...57B} {770, 57}

\bibitem[\protect\citeauthoryear{{Beifiori}, {Maraston}, {Thomas}  \&
  {Johansson}}{{Beifiori} et~al.}{2011}]{Beifiori2011}
{Beifiori} A.,  {Maraston} C.,  {Thomas} D.,   {Johansson} J.,  2011, \mn@doi
  [\aap] {10.1051/0004-6361/201016323}, \href
  {http://adsabs.harvard.edu/abs/2011A%26A...531A.109B} {531, A109}

\bibitem[\protect\citeauthoryear{{Belli}, {Newman}  \& {Ellis}}{{Belli}
  et~al.}{2014}]{Belli2014a}
{Belli} S.,  {Newman} A.~B.,   {Ellis} R.~S.,  2014, \mn@doi [\apj]
  {10.1088/0004-637X/783/2/117}, \href
  {http://adsabs.harvard.edu/abs/2014ApJ...783..117B} {783, 117}

\bibitem[\protect\citeauthoryear{{Bigiel}, {Leroy}, {Walter}, {Brinks}, {de
  Blok}, {Madore}  \& {Thornley}}{{Bigiel} et~al.}{2008}]{Bigiel2008}
{Bigiel} F.,  {Leroy} A.,  {Walter} F.,  {Brinks} E.,  {de Blok} W.~J.~G.,
  {Madore} B.,   {Thornley} M.~D.,  2008, \mn@doi [\aj]
  {10.1088/0004-6256/136/6/2846}, \href
  {http://adsabs.harvard.edu/abs/2008AJ....136.2846B} {136, 2846}

\bibitem[\protect\citeauthoryear{{Blanc} et~al.,}{{Blanc}
  et~al.}{2013}]{Blanc2013}
{Blanc} G.~A.,  et~al., 2013, \mn@doi [\aj] {10.1088/0004-6256/145/5/138},
  \href {http://adsabs.harvard.edu/abs/2013AJ....145..138B} {145, 138}

\bibitem[\protect\citeauthoryear{{Burkert} et~al.,}{{Burkert}
  et~al.}{2015}]{Burkert2015}
{Burkert} A.,  et~al., 2015, preprint, \href
  {http://adsabs.harvard.edu/abs/2015arXiv151003262B} {} (\mn@eprint {arXiv}
  {1510.03262})

\bibitem[\protect\citeauthoryear{{Cappellari}}{{Cappellari}}{2012}]{ppxf_software}
{Cappellari} M.,  2012, {pPXF: Penalized Pixel-Fitting stellar kinematics
  extraction} (\mn@eprint {ascl} {1210.002})

\bibitem[\protect\citeauthoryear{{Cappellari} \& {Copin}}{{Cappellari} \&
  {Copin}}{2003}]{Cappellari2003}
{Cappellari} M.,  {Copin} Y.,  2003, \mn@doi [\mnras]
  {10.1046/j.1365-8711.2003.06541.x}, \href
  {http://adsabs.harvard.edu/abs/2003MNRAS.342..345C} {342, 345}

\bibitem[\protect\citeauthoryear{{Cappellari} \& {Emsellem}}{{Cappellari} \&
  {Emsellem}}{2004}]{Cappellari2004}
{Cappellari} M.,  {Emsellem} E.,  2004, \mn@doi [\pasp] {10.1086/381875}, \href
  {http://adsabs.harvard.edu/abs/2004PASP..116..138C} {116, 138}

\bibitem[\protect\citeauthoryear{{Cappellari}, {Emsellem}, {Krajnovi{\'c}}
  et~al.}{{Cappellari} et~al.}{2011}]{Cappellari2011}
{Cappellari} M.,  {Emsellem} E.,  {Krajnovi{\'c}} D.,   et~al., 2011, \mn@doi
  [\mnras] {10.1111/j.1365-2966.2010.18174.x}, \href
  {http://adsabs.harvard.edu/abs/2011MNRAS.413..813C} {413, 813}

\bibitem[\protect\citeauthoryear{{Cresci} et~al.,}{{Cresci}
  et~al.}{2009}]{Cresci2009}
{Cresci} G.,  et~al., 2009, \mn@doi [\apj] {10.1088/0004-637X/697/1/115}, \href
  {http://adsabs.harvard.edu/abs/2009ApJ...697..115C} {697, 115}

\bibitem[\protect\citeauthoryear{{D'Eugenio}, {Houghton}, {Davies}  \& {Dalla
  Bont{\`a}}}{{D'Eugenio} et~al.}{2013}]{Deugenio2013}
{D'Eugenio} F.,  {Houghton} R.~C.~W.,  {Davies} R.~L.,   {Dalla Bont{\`a}} E.,
  2013, \mn@doi [\mnras] {10.1093/mnras/sts406}, \href
  {http://adsabs.harvard.edu/abs/2013MNRAS.429.1258D} {429, 1258}

\bibitem[\protect\citeauthoryear{{Davies} et~al.,}{{Davies}
  et~al.}{2011}]{Davies2011}
{Davies} R.,  et~al., 2011, \mn@doi [\apj] {10.1088/0004-637X/741/2/69}, \href
  {http://adsabs.harvard.edu/abs/2011ApJ...741...69D} {741, 69}

\bibitem[\protect\citeauthoryear{{Dubois} \& {Teyssier}}{{Dubois} \&
  {Teyssier}}{2008}]{Dubois2008}
{Dubois} Y.,  {Teyssier} R.,  2008, \mn@doi [\aap]
  {10.1051/0004-6361:20078326}, \href
  {http://adsabs.harvard.edu/abs/2008A%26A...477...79D} {477, 79}

\bibitem[\protect\citeauthoryear{{Dunkley} et~al.,}{{Dunkley}
  et~al.}{2009}]{Dunkley2009}
{Dunkley} J.,  et~al., 2009, \mn@doi [\apjs] {10.1088/0067-0049/180/2/306},
  \href {http://adsabs.harvard.edu/abs/2009ApJS..180..306D} {180, 306}

\bibitem[\protect\citeauthoryear{{F{\"o}rster Schreiber}, {Genzel}, {Bouche}
  et~al.}{{F{\"o}rster Schreiber} et~al.}{2009}]{ForsterSchreiber2009}
{F{\"o}rster Schreiber} N.~M.,  {Genzel} R.,  {Bouche} N.,   et~al., 2009,
  \mn@doi [\apj] {10.1088/0004-637X/706/2/1364}, \href
  {http://adsabs.harvard.edu/abs/2009ApJ...706.1364F} {706, 1364}

\bibitem[\protect\citeauthoryear{{Green} et~al.,}{{Green}
  et~al.}{2010}]{Green2010}
{Green} A.~W.,  et~al., 2010, \mn@doi [\nat] {10.1038/nature09452}, \href
  {http://adsabs.harvard.edu/abs/2010Natur.467..684G} {467, 684}

\bibitem[\protect\citeauthoryear{{Greene} \& {Ho}}{{Greene} \&
  {Ho}}{2006}]{Greene2006}
{Greene} J.~E.,  {Ho} L.~C.,  2006, \mn@doi [\apj] {10.1086/500353}, \href
  {http://adsabs.harvard.edu/abs/2006ApJ...641..117G} {641, 117}

\bibitem[\protect\citeauthoryear{{Hopkins}, {Quataert}  \& {Murray}}{{Hopkins}
  et~al.}{2011}]{Hopkins2011}
{Hopkins} P.~F.,  {Quataert} E.,   {Murray} N.,  2011, \mn@doi [\mnras]
  {10.1111/j.1365-2966.2011.19306.x}, \href
  {http://adsabs.harvard.edu/abs/2011MNRAS.417..950H} {417, 950}

\bibitem[\protect\citeauthoryear{{Horne}}{{Horne}}{1986}]{Horne1986}
{Horne} K.,  1986, \mn@doi [\pasp] {10.1086/131801}, \href
  {http://adsabs.harvard.edu/abs/1986PASP...98..609H} {98, 609}

\bibitem[\protect\citeauthoryear{{Ilbert}, {McCracken}, {Le F{\`e}vre},
  {Capak}, {Dunlop}  et~al.}{{Ilbert} et~al.}{2013}]{Ilbert2013}
{Ilbert} O.,  {McCracken} H.~J.,  {Le F{\`e}vre} O.,  {Capak} P.,  {Dunlop} J.,
    et~al., 2013, \mn@doi [\aap] {10.1051/0004-6361/201321100}, \href
  {http://adsabs.harvard.edu/abs/2013A%26A...556A..55I} {556, A55}

\bibitem[\protect\citeauthoryear{Ivezi\'c et~al.,}{Ivezi\'c
  et~al.}{2008}]{Ivezic2008}
Ivezi\'c v.,  et~al., 2008

\bibitem[\protect\citeauthoryear{{Katz}, {Weinberg}  \& {Hernquist}}{{Katz}
  et~al.}{1996}]{Katz1996}
{Katz} N.,  {Weinberg} D.~H.,   {Hernquist} L.,  1996, \mn@doi [\apjs]
  {10.1086/192305}, \href {http://adsabs.harvard.edu/abs/1996ApJS..105...19K}
  {105, 19}

\bibitem[\protect\citeauthoryear{{Kimm}, {Devriendt}, {Slyz}, {Pichon},
  {Kassin}  \& {Dubois}}{{Kimm} et~al.}{2011}]{Kimm2011}
{Kimm} T.,  {Devriendt} J.,  {Slyz} A.,  {Pichon} C.,  {Kassin} S.~A.,
  {Dubois} Y.,  2011, preprint, \href
  {http://adsabs.harvard.edu/abs/2011arXiv1106.0538K} {} (\mn@eprint {arXiv}
  {1106.0538})

\bibitem[\protect\citeauthoryear{{Kimm}, {Cen}, {Devriendt}, {Dubois}  \&
  {Slyz}}{{Kimm} et~al.}{2015}]{Kimm2015}
{Kimm} T.,  {Cen} R.,  {Devriendt} J.,  {Dubois} Y.,   {Slyz} A.,  2015,
  \mn@doi [\mnras] {10.1093/mnras/stv1211}, \href
  {http://adsabs.harvard.edu/abs/2015MNRAS.451.2900K} {451, 2900}

\bibitem[\protect\citeauthoryear{{Kormendy} \& {Illingworth}}{{Kormendy} \&
  {Illingworth}}{1982}]{Kormendy1982}
{Kormendy} J.,  {Illingworth} G.,  1982, \mn@doi [\apj] {10.1086/159923}, \href
  {http://adsabs.harvard.edu/abs/1982ApJ...256..460K} {256, 460}

\bibitem[\protect\citeauthoryear{{LSST Science Collaboration} et~al.,}{{LSST
  Science Collaboration} et~al.}{2009}]{LSST_book}
{LSST Science Collaboration} et~al., 2009, preprint, \href
  {http://adsabs.harvard.edu/abs/2009arXiv0912.0201L} {} (\mn@eprint {arXiv}
  {0912.0201})

\bibitem[\protect\citeauthoryear{{Madau} \& {Dickinson}}{{Madau} \&
  {Dickinson}}{2014}]{Madau2014}
{Madau} P.,  {Dickinson} M.,  2014, \mn@doi [\araa]
  {10.1146/annurev-astro-081811-125615}, \href
  {http://adsabs.harvard.edu/abs/2014ARA%26A..52..415M} {52, 415}

\bibitem[\protect\citeauthoryear{{Maraston}}{{Maraston}}{2005}]{Maraston2005}
{Maraston} C.,  2005, \mn@doi [\mnras] {10.1111/j.1365-2966.2005.09270.x},
  \href {http://adsabs.harvard.edu/abs/2005MNRAS.362..799M} {362, 799}

\bibitem[\protect\citeauthoryear{{Marchesini}, {van Dokkum}, {F{\"o}rster
  Schreiber}, {Franx}, {Labb{\'e}}  \& {Wuyts}}{{Marchesini}
  et~al.}{2009}]{Marchesini2009}
{Marchesini} D.,  {van Dokkum} P.~G.,  {F{\"o}rster Schreiber} N.~M.,  {Franx}
  M.,  {Labb{\'e}} I.,   {Wuyts} S.,  2009, \mn@doi [\apj]
  {10.1088/0004-637X/701/2/1765}, \href
  {http://adsabs.harvard.edu/abs/2009ApJ...701.1765M} {701, 1765}

\bibitem[\protect\citeauthoryear{{Marchesini} et~al.,}{{Marchesini}
  et~al.}{2010}]{Marchesini2010}
{Marchesini} D.,  et~al., 2010, \mn@doi [\apj] {10.1088/0004-637X/725/1/1277},
  \href {http://adsabs.harvard.edu/abs/2010ApJ...725.1277M} {725, 1277}

\bibitem[\protect\citeauthoryear{{Marchesini}, {Stefanon}, {Brammer}  \&
  {Whitaker}}{{Marchesini} et~al.}{2012}]{Marchesini2012}
{Marchesini} D.,  {Stefanon} M.,  {Brammer} G.~B.,   {Whitaker} K.~E.,  2012,
  \mn@doi [\apj] {10.1088/0004-637X/748/2/126}, \href
  {http://adsabs.harvard.edu/abs/2012ApJ...748..126M} {748, 126}

\bibitem[\protect\citeauthoryear{{Marsan} et~al.,}{{Marsan}
  et~al.}{2015}]{Marsan2015}
{Marsan} Z.~C.,  et~al., 2015, \mn@doi [\apj] {10.1088/0004-637X/801/2/133},
  \href {http://adsabs.harvard.edu/abs/2015ApJ...801..133M} {801, 133}

\bibitem[\protect\citeauthoryear{{Mendel} et~al.,}{{Mendel}
  et~al.}{2015}]{Mendel2015}
{Mendel} J.~T.,  et~al., 2015, \mn@doi [\apjl] {10.1088/2041-8205/804/1/L4},
  \href {http://adsabs.harvard.edu/abs/2015ApJ...804L...4M} {804, L4}

\bibitem[\protect\citeauthoryear{{Naylor}}{{Naylor}}{1998}]{Naylor1998}
{Naylor} T.,  1998, \mn@doi [\mnras] {10.1046/j.1365-8711.1998.01314.x}, \href
  {http://adsabs.harvard.edu/abs/1998MNRAS.296..339N} {296, 339}

\bibitem[\protect\citeauthoryear{{Peng}, {Ho}, {Impey}  \& {Rix}}{{Peng}
  et~al.}{2002}]{Peng2002}
{Peng} C.~Y.,  {Ho} L.~C.,  {Impey} C.~D.,   {Rix} H.-W.,  2002, \mn@doi [\aj]
  {10.1086/340952}, \href {http://adsabs.harvard.edu/abs/2002AJ....124..266P}
  {124, 266}

\bibitem[\protect\citeauthoryear{{Peng}, {Ho}, {Impey}  \& {Rix}}{{Peng}
  et~al.}{2010}]{Peng2010}
{Peng} C.~Y.,  {Ho} L.~C.,  {Impey} C.~D.,   {Rix} H.-W.,  2010, \mn@doi [\aj]
  {10.1088/0004-6256/139/6/2097}, \href
  {http://adsabs.harvard.edu/abs/2010AJ....139.2097P} {139, 2097}

\bibitem[\protect\citeauthoryear{{Pontzen}, {Ro{\v s}kar}, {Stinson}  \&
  {Woods}}{{Pontzen} et~al.}{2013}]{pynbody}
{Pontzen} A.,  {Ro{\v s}kar} R.,  {Stinson} G.,   {Woods} R.,  2013, {pynbody:
  N-Body/SPH analysis for python} (\mn@eprint {ascl} {1305.002})

\bibitem[\protect\citeauthoryear{{Powell}, {Slyz}  \& {Devriendt}}{{Powell}
  et~al.}{2011}]{Powell2011}
{Powell} L.~C.,  {Slyz} A.,   {Devriendt} J.,  2011, \mn@doi [\mnras]
  {10.1111/j.1365-2966.2011.18668.x}, \href
  {http://adsabs.harvard.edu/abs/2011MNRAS.414.3671P} {414, 3671}

\bibitem[\protect\citeauthoryear{{Rodr{\'{\i}}guez Del Pino}, {Bamford},
  {Arag{\'o}n-Salamanca}, {Milvang-Jensen}, {Merrifield}  \&
  {Balcells}}{{Rodr{\'{\i}}guez Del Pino} et~al.}{2014}]{Rodriguez2014}
{Rodr{\'{\i}}guez Del Pino} B.,  {Bamford} S.~P.,  {Arag{\'o}n-Salamanca} A.,
  {Milvang-Jensen} B.,  {Merrifield} M.~R.,   {Balcells} M.,  2014, \mn@doi
  [\mnras] {10.1093/mnras/stt2202}, \href
  {http://adsabs.harvard.edu/abs/2014MNRAS.438.1038R} {438, 1038}

\bibitem[\protect\citeauthoryear{{Salpeter}}{{Salpeter}}{1955}]{Salpeter1955}
{Salpeter} E.~E.,  1955, \mn@doi [\apj] {10.1086/145971}, \href
  {http://adsabs.harvard.edu/abs/1955ApJ...121..161S} {121, 161}

\bibitem[\protect\citeauthoryear{{Sanchez}, {Kennicutt}, {Gil de Paz}
  et~al.}{{Sanchez} et~al.}{2012}]{Sanchez2012}
{Sanchez} S.~F.,  {Kennicutt} R.~C.,  {Gil de Paz} A.,   et~al., 2012, \mn@doi
  [\aap] {10.1051/0004-6361/201117353}, \href
  {http://adsabs.harvard.edu/abs/2012A%26A...538A...8S} {538, A8}

\bibitem[\protect\citeauthoryear{{Steidel} et~al.,}{{Steidel}
  et~al.}{2014}]{Steidel2014}
{Steidel} C.~C.,  et~al., 2014, \mn@doi [\apj] {10.1088/0004-637X/795/2/165},
  \href {http://adsabs.harvard.edu/abs/2014ApJ...795..165S} {795, 165}

\bibitem[\protect\citeauthoryear{{Straatman}, {Labb{\'e}}, {Spitler}, {Allen},
  {Altieri}  et~al.}{{Straatman} et~al.}{2014}]{Straatman2014}
{Straatman} C.~M.~S.,  {Labb{\'e}} I.,  {Spitler} L.~R.,  {Allen} R.,
  {Altieri} B.,   et~al., 2014, \mn@doi [\apjl] {10.1088/2041-8205/783/1/L14},
  \href {http://adsabs.harvard.edu/abs/2014ApJ...783L..14S} {783, L14}

\bibitem[\protect\citeauthoryear{{Sutherland} \& {Dopita}}{{Sutherland} \&
  {Dopita}}{1993}]{Sutherland1993}
{Sutherland} R.~S.,  {Dopita} M.~A.,  1993, \mn@doi [\apjs] {10.1086/191823},
  \href {http://adsabs.harvard.edu/abs/1993ApJS...88..253S} {88, 253}

\bibitem[\protect\citeauthoryear{{Teyssier}}{{Teyssier}}{2002}]{Teyssier2002}
{Teyssier} R.,  2002, \mn@doi [\aap] {10.1051/0004-6361:20011817}, \href
  {http://adsabs.harvard.edu/abs/2002A%26A...385..337T} {385, 337}

\bibitem[\protect\citeauthoryear{{Thatte}, {Scott}, {Houghton}, {Nuernberger},
  {Abuter}  \& {Tecza}}{{Thatte} et~al.}{2012}]{Thatte2012}
{Thatte} N.~A.,  {Scott} N.,  {Houghton} R.,  {Nuernberger} D.,  {Abuter}
  R.~N.,   {Tecza} M.,  2012, in Observatory Operations: Strategies, Processes,
  and Systems IV. p. 844809, \mn@doi{10.1117/12.926145}

\bibitem[\protect\citeauthoryear{{Thatte}, {Clarke}, {Bryson}  et~al.}{{Thatte}
  et~al.}{2014}]{Thatte2014}
{Thatte} N.~A.,  {Clarke} F.,  {Bryson} I.,   et~al., 2014, in Society of
  Photo-Optical Instrumentation Engineers (SPIE) Conference Series. p.~25,
  \mn@doi{10.1117/12.2055436}

\bibitem[\protect\citeauthoryear{{Thomas}, {Maraston}, {Bender}  \& {Mendes de
  Oliveira}}{{Thomas} et~al.}{2005}]{Thomas2005}
{Thomas} D.,  {Maraston} C.,  {Bender} R.,   {Mendes de Oliveira} C.,  2005,
  \mn@doi [\apj] {10.1086/426932}, \href
  {http://adsabs.harvard.edu/abs/2005ApJ...621..673T} {621, 673}

\bibitem[\protect\citeauthoryear{{Tonini}, {Maraston}, {Devriendt}, {Thomas}
  \& {Silk}}{{Tonini} et~al.}{2009}]{Tonini2009}
{Tonini} C.,  {Maraston} C.,  {Devriendt} J.,  {Thomas} D.,   {Silk} J.,  2009,
  \mn@doi [\mnras] {10.1111/j.1745-3933.2009.00657.x}, \href
  {http://adsabs.harvard.edu/abs/2009MNRAS.396L..36T} {396, L36}

\bibitem[\protect\citeauthoryear{{Vazdekis}, {Casuso}, {Peletier}  \&
  {Beckman}}{{Vazdekis} et~al.}{1996}]{Vazdekis1996}
{Vazdekis} A.,  {Casuso} E.,  {Peletier} R.~F.,   {Beckman} J.~E.,  1996,
  \mn@doi [\apjs] {10.1086/192340}, \href
  {http://adsabs.harvard.edu/abs/1996ApJS..106..307V} {106, 307}

\bibitem[\protect\citeauthoryear{{Vazdekis}, {S{\'a}nchez-Bl{\'a}zquez},
  {Falc{\'o}n-Barroso}, {Cenarro}, {Beasley}, {Cardiel}, {Gorgas}  \&
  {Peletier}}{{Vazdekis} et~al.}{2010}]{Vazdekis2010}
{Vazdekis} A.,  {S{\'a}nchez-Bl{\'a}zquez} P.,  {Falc{\'o}n-Barroso} J.,
  {Cenarro} A.~J.,  {Beasley} M.~A.,  {Cardiel} N.,  {Gorgas} J.,   {Peletier}
  R.~F.,  2010, \mn@doi [\mnras] {10.1111/j.1365-2966.2010.16407.x}, \href
  {http://adsabs.harvard.edu/abs/2010MNRAS.404.1639V} {404, 1639}

\bibitem[\protect\citeauthoryear{{Vazdekis}, {Ricciardelli}, {Cenarro},
  {Rivero-Gonz{\'a}lez}, {D{\'{\i}}az-Garc{\'{\i}}a}  \&
  {Falc{\'o}n-Barroso}}{{Vazdekis} et~al.}{2012}]{Vazdekis2012}
{Vazdekis} A.,  {Ricciardelli} E.,  {Cenarro} A.~J.,  {Rivero-Gonz{\'a}lez}
  J.~G.,  {D{\'{\i}}az-Garc{\'{\i}}a} L.~A.,   {Falc{\'o}n-Barroso} J.,  2012,
  \mn@doi [\mnras] {10.1111/j.1365-2966.2012.21179.x}, \href
  {http://adsabs.harvard.edu/abs/2012MNRAS.424..157V} {424, 157}

\bibitem[\protect\citeauthoryear{{Wisnioski}, {F{\"o}rster Schreiber}, {Wuyts}
  et~al.}{{Wisnioski} et~al.}{2014}]{Wisnioski2014}
{Wisnioski} E.,  {F{\"o}rster Schreiber} N.~M.,  {Wuyts} S.,   et~al., 2014,
  preprint, \href {http://adsabs.harvard.edu/abs/2014arXiv1409.6791W} {}
  (\mn@eprint {arXiv} {1409.6791})

\bibitem[\protect\citeauthoryear{{Zieleniewski} \& {Thatte}}{{Zieleniewski} \&
  {Thatte}}{2013}]{Zieleniewski2013}
{Zieleniewski} S.,  {Thatte} N.,  2013, in {Esposito} S.,  {Fini} L.,  eds,
  Proceedings of the Third AO4ELT Conference. p.~43,
  \mn@doi{10.12839/AO4ELT3.13269}

\bibitem[\protect\citeauthoryear{Zieleniewski, Thatte, Kendrew, Houghton,
  Tecza, Clarke, Fusco  \& Swinbank}{Zieleniewski
  et~al.}{2014}]{Zieleniewski2014}
Zieleniewski S.,  Thatte N.,  Kendrew S.,  Houghton R.,  Tecza M.,  Clarke F.,
  Fusco T.,   Swinbank M.,  2014. pp 914793--914793--12,
  \mn@doi{10.1117/12.2055578}, \url {http://dx.doi.org/10.1117/12.2055578}

\bibitem[\protect\citeauthoryear{{Zieleniewski}, {Thatte}, {Kendrew},
  {Houghton}, {Swinbank}, {Tecza}, {Clarke}  \& {Fusco}}{{Zieleniewski}
  et~al.}{2015}]{Zieleniewski2015}
{Zieleniewski} S.,  {Thatte} N.,  {Kendrew} S.,  {Houghton} R.~C.~W.,
  {Swinbank} A.~M.,  {Tecza} M.,  {Clarke} F.,   {Fusco} T.,  2015, preprint,
  \href {http://adsabs.harvard.edu/abs/2015arXiv150804441Z} {} (\mn@eprint
  {arXiv} {1508.04441})

\bibitem[\protect\citeauthoryear{{van de Sande} et~al.,}{{van de Sande}
  et~al.}{2013}]{vdSande2013}
{van de Sande} J.,  et~al., 2013, \mn@doi [\apj] {10.1088/0004-637X/771/2/85},
  \href {http://adsabs.harvard.edu/abs/2013ApJ...771...85V} {771, 85}

\bibitem[\protect\citeauthoryear{{van der Wel}, {Franx}, {van Dokkum},
  {Skelton}, {Momcheva}, {Whitaker}, {Brammer}  et~al.}{{van der Wel}
  et~al.}{2014}]{vanderWel2014}
{van der Wel} A.,  {Franx} M.,  {van Dokkum} P.~G.,  {Skelton} R.~E.,
  {Momcheva} I.~G.,  {Whitaker} K.~E.,  {Brammer} G.~B.,   et~al., 2014,
  \mn@doi [\apj] {10.1088/0004-637X/788/1/28}, \href
  {http://adsabs.harvard.edu/abs/2014ApJ...788...28V} {788, 28}

\makeatother
\end{thebibliography}

\label{lastpage}

\end{document}